\documentclass{turing2012}

\usepackage{array,multirow,graphicx}
\usepackage{amsmath,amsthm,amssymb}
\usepackage{cleveref}
\usepackage{overpic}

\theoremstyle{definition}
\newtheorem{definition}{Definition}
\newtheorem{example}{Example}

\begin{document}

\title{A Model of Multi-Agent Consensus for Vague and Uncertain Beliefs}

\author{
Michael Crosscombe \and Jonathan Lawry\\
Department of Engineering Mathematics, \\University of Bristol, \\BS8 1UB, United Kingdom\\\
\texttt{m.crosscombe@bristol.ac.uk $\cdot$ j.lawry@bristol.ac.uk}
}

\maketitle

\begin{abstract}
Consensus formation is investigated for multi-agent systems in which agents' beliefs are both vague and uncertain. Vagueness is represented by a third truth state meaning \emph{borderline}. This is combined with a probabilistic model of uncertainty. A belief combination operator is then proposed which exploits borderline truth values to enable agents with conflicting beliefs to reach a compromise. A number of simulation experiments are carried out in which agents apply this operator in pairwise interactions, under the bounded confidence restriction that the two agents' beliefs must be sufficiently consistent with each other before agreement can be reached. As well as studying the consensus operator in isolation we also investigate scenarios in which agents are influenced either directly or indirectly by the state of the world. For the former we conduct simulations which combine consensus formation with belief updating based on evidence. For the latter we investigate the effect of assuming that the closer an agent's beliefs are to the truth the more visible they are in the consensus building process. In all cases applying the consensus operators results in the population converging to a single shared belief which is both crisp and certain. Furthermore, simulations which combine consensus formation with evidential updating converge faster to a shared opinion which is closer to the actual state of the world than those in which beliefs are only changed as a result of directly receiving new evidence. Finally, if agent interactions are guided by belief quality measured as similarity to the true state of the world, then applying the consensus operator alone results in the population converging to a high quality shared belief.

\end{abstract}

\hspace{1em}


\section{Introduction}
Reaching an agreement by identifying a position or viewpoint which can ultimately be accepted by a significant proportion of the individuals in a population is a fundamental part of many multi-agent decision making and negotiation scenarios. In human interactions opinions can take the form of vague propositions with explicitly borderline truth values i.e. where the proposition is neither absolutely true nor absolutely false~\cite{keefe}. Indeed a number of recent studies~\cite{crosscombe,balenzuela,perron,lama,vazquez} have suggested that the presence of an intermediate truth state of this kind can play a positive role in opinion dynamics by allowing compromise and hence facilitating convergence to a shared viewpoint.

In addition to vagueness, individuals often have uncertain beliefs due to the limited and imperfect evidence that they have available to them about the true state of the world. In this paper we propose a model of belief combination by which two independent agents can reach a consensus between distinct and, to some extent, conflicting opinions which are both uncertain and vague. We show that in an agent-based system, iteratively applying this operator under a variety of conditions results in the agents converging on a single opinion which is both crisp (i.e. non-vague) and certain.

However, beliefs are not arrived at only as the result of consensus building within a closed system, but are also influenced by the actual state of the world. This can be both by agents updating their beliefs given evidence, and by them receiving different levels of payoff for decisions and actions taken on the basis  of their beliefs. In this paper we model both of these processes when combined with consensus formation. We consider the case in which a population of agents interact continually at random, forming consensus where appropriate, but occasionally receiving direct information about the state of the world. Defining a measure of belief quality taking account of the similarity between an agent's belief and the true state of the world, we then record this quality measure in simulations which combine both consensus building and belief updating from evidence and compare these with simulations in which only evidence based updating occurs. In these studies we observe that combining evidence based updating and consensus building results in faster convergence to higher quality beliefs than when beliefs are only changed as a result of receiving new evidence. This would seem to offer some support for the hypothesis put forward in~\cite{douvenkelp}, that scientists may gain by taking account of each others opinions as well as by considering direct evidence.

In addition to direct evidence there are also indirect mechanisms by which agents receive feedback on the quality of beliefs. For example,  when an agent makes decisions and takes actions based on their beliefs they may receive some form of reward or payoff. In such cases it is reasonable to assume that the higher the quality of an agent's beliefs, i.e. the closer they are to the true state of the world, the higher the payoff that the agent will receive on average. Here we investigate a scenario in which the quality of an agent's beliefs influences their visibility in the consensus building process. This is studied in simulation experiments in which interactions between agents are guided by the quality of their beliefs, so that individuals holding higher quality opinions are more likely to be selected to combine their beliefs.

An outline of the remainder of the paper is as follows: Section 2 gives an overview of related work in this area. Section 3 introduces a propositional model of belief which incorporates both vagueness and uncertainty, and proposes a combination operator for generating a compromise between two distinct beliefs. In section 4 a set of simulation experiments are described in which agents interact at random and apply the operator introduced in section 3 provided that they hold sufficiently consistent beliefs. Section 5 combines random agent interactions and consensus formation with belief updating based on direct evidence about the true state of the world. In section 6 we describe simulation experiments in which agent interactions are dependent on the quality of their beliefs. Finally in section 7 we present some conclusions and discuss possible future directions.

\section{Background and Related Work}
A number of studies in the opinion dynamics literature exploit a third truth state to aid convergence and also to mitigate the effect of a minority of highly opinionated individuals. For example,~\cite{lama} and~\cite{vazquez} study scenarios in which interactions only take place between agents with a clear viewpoint and undecided agents. Alternatively, in~\cite{balenzuela} the three truth states are defined by applying a partitioning threshold to an underlying real value. Updating is pairwise between agents and takes place incrementally on the real values, but where the magnitude and sign of the increments depends on the current truth states of the agents involved. An alternative pairwise three-valued operator is proposed in~\cite{perron} which is applied directly to truth states. In particular, this operator assigns the third truth state as a compromise between two opinions with strictly opposing truth values. The logical properties of this operator and its relationship to other similar aggregation functions are investigated in~\cite{lawrydubois}. For a language with a single proposition and assuming unconstrained random interactions between individuals~\cite{perron} proves convergence to a single shared Boolean opinion. This framework is extended in~\cite{crosscombe} to languages with multiple propositions and to include a form of bounded confidence (see~\cite{hegselmann02}) in which interactions only take place between individuals with sufficiently consistent opinions. Furthermore,~\cite{crosscombe} also investigates convergence when the selection of agents is guided by a measure of the quality of their opinions and shows that the average quality of opinions across the population is higher at steady state than at initialisation.

One common feature of most of the studies mentioned above is that, either explicitly or implicitly, they interpret the third truth value as meaning `uncertain' or `unknown'. In contrast, as stated in section 1, we intend the middle truth value to refer to borderline cases resulting from the underlying vagueness of the language. So, for example, giving the proposition `Ethel is short' the intermediate truth value means that Ethel's height is borderline short/not short, rather than meaning that Ethel's height is unknown. This approach allows us to distinguish between vagueness and uncertainty so that, for instance, based on their knowledge of Ethel's height an agent could be certain that she is borderline short. A more detailed analysis of the difference between these two possible interpretations of the third truth state is given in~\cite{ciucci}.

The idea of bounded confidence~\cite{hegselmann02,deffuant02} has been proposed as a mechanism by which agents limit their interactions with others, so that they only combine their beliefs with individuals holding opinions which are sufficiently similar to their current view. A version of bounded confidence is also used in our proposed model, where each agent measures the relative inconsistency of their beliefs with those of others, and is then only willing to combine beliefs with agents whose inconsistency measure is below a certain threshold.

The aggregation of uncertain beliefs in the form of a probability distribution over some underlying parameter has been widely studied with work on opinion pooling dating back to~\cite{stone} and~\cite{degroot}. Usually the aggregate of a set of opinions takes the form of a weighted linear combination of the associated probability distributions. However, the convergence of alternative opinion pooling functions has been studied in~\cite{hegselmann05} and axiomatic characterisations of different operators are given in~\cite{dietrich}. All of these approaches assume Boolean truth states and indeed there are very few studies in this context which combine probability with a three-valued truth model. One such is~\cite{cho} which adopts a model of beliefs in the form of Dempster-Shafer functions. The combination operators proposed in~\cite{cho}, however, are quite different to those described in the current paper and result in quite different limiting behaviour. The operator investigated below was first proposed in~\cite{lawrydubois} as an extension of the approach of~\cite{perron} so as to also take account of probabilistic uncertainty and to our knowledge it has not, up to this point, been studied in an agent-based setting. Hence, in contrast to~\cite{lawrydubois} the focus of this current paper is on the system level behaviour of the proposed operator rather than on the theoretical properties.

\section{A Consensus Operator for Vague and Uncertain Beliefs}
We consider a simple language consisting of $n$ propositions ${\cal L}=\{p_1, \ldots, p_n\}$. Each can have one of the three truth values $0$, denoting false, $\frac{1}{2}$, denoting borderline and $1$, denoting true. A \emph{valuation} of ${\cal  L}$ corresponds to an allocation of a truth value to each of the propositions. Consequently, a valuation is naturally represented as an $n$ dimensional vector $\mathbf{v} \in \{0,\frac{1}{2},1\}^n$. We let $\mathbf{v}(p_i)$ denote the $i$'th dimension of $\mathbf{v}$ as corresponding to the truth value of the proposition $p_i$ in the valuation $\mathbf{v}$. In the absence of any uncertainty we assume than an agent's opinion is represented by a single valuation. For two agents with distinct and possibly conflicting opinions $\mathbf{v}_1, \mathbf{v}_2 \in \{0,\frac{1}{2},1\}^n$ to reach a compromise position or consensus we propose an operator introduced in~\cite{perron} and~\cite{lawrydubois}, and based on the truth table given in Table~\ref{tab:consensus_op} which is applied to each proposition independently so that:
\begin{gather*}
	\mathbf{v}_1 \odot \mathbf{v}_2=(\mathbf{v}_1(p_1) \odot \mathbf{v}_2(p_1), \ldots,\mathbf{v}_1(p_n) \odot \mathbf{v}_2(p_n))
\end{gather*}
The intuition behind the operator is as follows: In the case that the two agents disagree then if one has allocated a non-borderline truth value to $p_i$, while the other has given $p_i$ a borderline truth value then the non-borderline truth value is adopted in the agreed compromise. In other words, if one agent has a strong view about $p_i$ while the other is ambivalent then they will both agree to adopt the strong viewpoint. In contrast if both agents have strong but opposing views i.e. with one valuation giving $p_i$ truth value $0$ and the other $1$, then they will agree on a compromise truth value of $\frac{1}{2}$.

\begin{table}
	\centering
	\begin{tabular}{| r || c | c | c | c |} 
		\hline 
		$\odot$& $1$ & $\frac{1}{2}$ & $0$  \\   [0.5ex]
		\hline \hline
		$1$ & $1$ & $1$ & $\frac{1}{2}$ \\ 
		\hline
		$\frac{1}{2}$ & $1$ & $\frac{1}{2}$ & $0$ \\
		\hline
		$0$ & $\frac{1}{2}$ & $0$ & $0$ \\ [1ex] 
		\hline
	\end{tabular}
	\vspace{1em}
	\caption{Truth Table for the Consensus Operator.}
	\label{tab:consensus_op}
\end{table}


Here we extend this model to allow agents to hold opinions which are uncertain as well as vague. More specifically, an integrated approach to uncertainty and vagueness is adopted in which an agent's belief is characterised by a probability distribution $w$ over $\{0,\frac{1}{2},1\}^n$ so that $w(\mathbf{v})$ quantifies the agent's belief that $\mathbf{v}$ is the correct valuation of ${\cal L}$. This naturally generates lower and upper belief measures on ${\cal L}$ quantifying the agent's belief that a given proposition is true and that it is not false respectively~\cite{lawrytang}. That is; for $p_i \in {\cal L}$, \footnote{In the following we slightly abuse notation and also use $w$ to denote the probability measure generated by the probability distribution $w$.}
\begin{gather*}
	\underline{\mu}(p_i) =w(\{\mathbf{v}: \mathbf{v}(p_i)=1 \}) \text{ and } \\ \overline{\mu}(p_i) =w(\{\mathbf{v}: \mathbf{v}(p_i) \neq 0\})
\end{gather*}
The probability of each of the possible truth values for a proposition $p_i$ can be recaptured from the lower and upper belief measures such that the probabilities that $p_i$ is true, borderline and false are given by $\underline{\mu}(p_i)$, $\overline{\mu}(p_i)-\underline{\mu}(p_i)$ and $1-\overline{\mu}(p_i)$ respectively. Hence, we can represent an agent's belief by a vector of pairs of lower and upper belief values for each proposition as follows:
\begin{gather*}
	\boldsymbol{\mu}=((\underline{\mu}(p_1),\overline{\mu}(p_1)),\ldots,(\underline{\mu}(p_n),\overline{\mu}(p_n)))
\end{gather*}
Here we let $\boldsymbol{\mu}(p_i)$ denote $(\underline{\mu}(p_i),\overline{\mu}(p_i))$, the pair of lower and upper belief values for $p_i$. In the case that a belief $\boldsymbol{\mu}$ gives probability zero to the borderline truth value for every proposition in ${\cal L}$ so that $\underline{\mu}(p_i)=\overline{\mu}(p_i)=\mu(p_i)$ for $i=1, \ldots,n$ then we call $\boldsymbol{\mu}$ a \emph{crisp belief}.

The following definition expands the consensus operation $\odot$ from three-valued valuations to this more general representation framework.

\begin{definition}{Consensus Operator for Belief Pairs}\\
	\label{conop}
	{\footnotesize
	\begin{gather*}
		\boldsymbol{\mu}_1 \odot \boldsymbol{\mu}_2=\\
	( (\underline{\mu}_1 \odot \underline{\mu}_2(p_1),\overline{\mu}_1 \odot \overline{\mu}_2(p_1)), \ldots,  (\underline{\mu}_1 \odot \underline{\mu}_2(p_n),\overline{\mu}_1 \odot \overline{\mu}_2(p_n))  )
	\end{gather*}
	}
	where
	\begin{gather*}
		\underline{\mu}_1 \odot \underline{\mu}_2(p_i)=\underline{\mu}_1(p_i) \times \overline{\mu}_2(p_i) + \overline{\mu}_1(p_i) \times \underline{\mu}_2(p_i) \\
	-\underline{\mu}_1(p_i) \times \underline{\mu}_2(p_i)
	\end{gather*}
	and
	\begin{gather*}
		\overline{\mu}_1 \odot \overline{\mu}_2(p_i)=\underline{\mu}_1(p_i) +  \underline{\mu}_2(p_i) + \overline{\mu}_1(p_i)\times \overline{\mu}_2(p_i) \\
	-\overline{\mu}_1(p_i)\times \underline{\mu}_2(p_i) - \underline{\mu}_1(p_i)\times \overline{\mu}_2(p_i)
	\end{gather*}
\end{definition}
If $\boldsymbol{\mu}_1$ and $\boldsymbol{\mu}_2$ are generated by the probability distributions $w_1$ and $w_2$ on $\{0,\frac{1}{2},1\}^n$ respectively, then $\boldsymbol{\mu}_1 \odot \boldsymbol{\mu}_2$ corresponds to the lower and upper measures generated by the following combined probability distribution on $\{0,\frac{1}{2},1\}^n$~\cite{lawrydubois}:
\begin{gather*}
	w_1 \odot w_2(\mathbf{v})=\sum_{\mathbf{v}_1, \mathbf{v}_2:\mathbf{v}_1 \odot \mathbf{v}_2=\mathbf{v}} w_1(\mathbf{v}_1) \times  w_2(\mathbf{v}_2)
\end{gather*}
In other words, assuming that the two agents are independent, all pairs of valuations supported by the two agents are combined using the consensus operator for valuations and then aggregated. Interestingly, this operator can be reformulated as a special case of the union combination operator in Dempster-Shafter theory (see~\cite{shafer}) proposed by~\cite{duboisprade88ds}. To see this notice that given a probability distribution $w$ on $\{0, \frac{1}{2},1 \}$ we can generate a Dempster-Shafer mass function $m$ on the power set of $\{0,1\}$ for each proposition $p_i$ such that:
\begin{gather*}
m(\{1\})=w(\{\mathbf{v}:\mathbf{v}(p_i)=1 \})=\underline{\mu}(p_i)\\
m(\{0\})=w(\{\mathbf{v}:\mathbf{v}(p_i)=0 \})=1-\overline{\mu}(p_i)\\
m(\{0,1\})=w(\{\mathbf{v}:\mathbf{v}(p_i)=\frac{1}{2} \})=\overline{\mu}(p_i)-\underline{\mu}(p_i)
\end{gather*}
In this reformulation then the lower and upper measures $\underline{\mu}(p_i)$ and $\overline{\mu}(p_i)$ correspond to the Dempster-Shafer belief and plausibility of $\{1\}$, as generated by $m$, respectively. Now in this context the union combination operator is defined as follows: Let $m_1$ and $m_2$ be two mass functions generated as above by probability distributions $w_1$ and $w_2$. Also let $c$ be a set combination function defined as:
\begin{gather*}
c(A,B)=\begin{cases} A \cap B: A \cap B \neq \emptyset \\ A \cup B: \text{otherwise} \end{cases}
\end{gather*}
Then the combination of $m_1$ and $m_2$ is defined by:
\begin{gather*}
m_1 \odot m_2 (D) = \sum_{A,B\subseteq \{0,1\}: c(A,B)=D } m_1(A) \times m_2(B)
\end{gather*}
The belief and plausibility of $\{1\}$ generated by $m_1 \odot m_2$ then respectively correspond to $\underline{\mu}_1 \odot \underline{\mu}_2(p_i)$ and $\overline{\mu}_1 \odot \overline{\mu}_2(p_i)$ as given in Definition~\ref{conop}.
\begin{example}
Suppose two agents have the following beliefs about proposition $p_i$: $\boldsymbol{\mu}_1(p_i)=(0.6,0.8)$ and $\boldsymbol{\mu}_2(p_i)=(0.4,0.7)$. The associated probability distributions on valuations, $w_1$ and $w_2$, are then such that:
\begin{gather*}
w_1(\{\mathbf{v}:\mathbf{v}(p_i)=1\})=0.6,\\
w_1(\{\mathbf{v}:\mathbf{v}(p_i)=\frac{1}{2}\})=0.8-0.6=0.2,\\
w_1(\{\mathbf{v}:\mathbf{v}(p_i)=0\})=1-0.8=0.2 \text{ and}\\
w_2(\{\mathbf{v}:\mathbf{v}(p_i)=1\})=0.4,\\
w_2(\{\mathbf{v}:\mathbf{v}(p_i)=\frac{1}{2}\})=0.7-0.4=0.3,\\
w_2(\{\mathbf{v}:\mathbf{v}(p_i)=0\})=1-0.7=0.3
\end{gather*}
From this we can generate the probability table shown in Table~\ref{tab:prob_consensus_op}. Here the corresponding truth values are generated as in Table~\ref{tab:consensus_op} and the probability values in each cell are the product of the associated row and column probability values. From this table we can then determine the consensus belief in $p_i$ by taking the sum of the probabilities of the cells with truth value $1$ to give the lower measure and the sum of the probabilities of the cells with truth values of either $1$ or $\frac{1}{2}$ to give the upper measure. That is:
\begin{gather*}
\underline{\mu}_1 \odot \underline{\mu}_2(p_i)=0.24+0.08+0.18=0.5\\
\overline{\mu}_2 \odot \overline{\mu}_2(p_i)=0.24+0.08+0.18+ 0.18 +0.06+0.08\\=0.82
\end{gather*}

\begin{table}
	\centering
	\begin{tabular}{| r || c | c | c | c |} 
		\hline 
		$\odot$& $1:0.6$ & $\frac{1}{2}:0.2$ & $0:0.2$  \\   [0.5ex]
		\hline \hline
		$1:0.4$ & $1:0.24$ & $1:0.08$ & $\frac{1}{2}:0.08$ \\ 
		\hline
		$\frac{1}{2}:0.3$ & $1:0.18$ & $\frac{1}{2}:0.06$ & $0:0.06$ \\
		\hline
		$0:0.3$ & $\frac{1}{2}:0.18$ & $0:0.06$ & $0:0.06$ \\ [1ex] 
		\hline
	\end{tabular}
	\vspace{1em}
	\caption{Probability Table for the Consensus Operator.}
	\label{tab:prob_consensus_op}
\end{table}

\end{example}

We now introduce three measures which will subsequently be used to analyse the behaviour of multi-agent systems applying the operator given in Definition~\ref{conop}.
\begin{definition}{A Measure of Vagueness}\\
	\label{vagueness}
	The degree of vagueness of the belief $\boldsymbol{\mu}$ is given by:
	\begin{gather*}
		\frac{1}{n}\sum_{i=1}^n (\overline{\mu}(p_i)-\underline{\mu}(p_i))
	\end{gather*}
\end{definition}
Definition~\ref{vagueness} is simply the probability of the truth value $\frac{1}{2}$ averaged across the $n$ propositions in ${\cal L}$. Since in this model vagueness is associated with borderline truth values then this provides an intuitive measure of the degree of vagueness of an opinion. Accordingly the most vague belief has $(\underline{\mu}(p_i),\overline{\mu}(p_i))=(0,1)$ for $i=1, \ldots, n$.

\begin{definition}{A Measure of Uncertainty}\\
	\label{entropy}
	The entropy of the belief $\boldsymbol{\mu}$ is given by:
	\begin{gather*}
		\frac{1}{n}\sum_{i=1}^n H(p_i) \text{ where} \\
		H(p_i)=-\underline{\mu}(p_i)\log_2(\underline{\mu}(p_i))\\-(\overline{\mu}(p_i)
		-\underline{\mu}(p_i))\log_2(\overline{\mu}(p_i)-\underline{\mu}(p_i))\\
		-(1-\overline{\mu}(p_i))\log_2(1-\overline{\mu}(p_i))
	\end{gather*}
\end{definition}
Definition~\ref{entropy} corresponds to the entropy of the marginal distributions on $\{0,\frac{1}{2},1\}$ averaged across the $n$ propositions. Hence, according to this measure the most uncertain belief allocates probability $\frac{1}{3}$ to each of the truth values for each proposition so that:
\begin{gather*}
	\boldsymbol{\mu}=\left(\left(\frac{1}{3},\frac{2}{3}\right),\ldots, \left(\frac{1}{3},\frac{2}{3}\right)\right)
\end{gather*}
The most certain beliefs then corresponds to those for which for every proposition $(\underline{\mu}(p_i),\overline{\mu}(p_i))=(0,0)$, $(0,1)$ or $(1,1)$.
\begin{definition}{A Measure of Inconsistency}
	\\
	\label{inconsistency}
	The degree of inconsistency of two beliefs $\boldsymbol{\mu}_1$ and $\boldsymbol{\mu}_2$ is given by:
	\begin{gather*}
		\frac{1}{n}\sum_{i=1}^n \left( \underline{\mu}_1(p_i)\times(1-\overline{\mu}_2(p_i))+(1-\overline{\mu}_1(p_i))\times \underline{\mu}_2(p_i)\right)
	\end{gather*}
\end{definition}
Definition~\ref{inconsistency} is the probability of a direct conflict between the two agents' beliefs, i.e. with agent $1$ allocating the truth value $1$ and agent $2$ the truth value $0$ or vice versa, this being then averaged across all $n$ propositions.

\section{Simulation Experiments with Random Selection of Agents}

We now describe simulation experiments in which pairs of agents are selected to interact at random. A model of bounded confidence is applied according to which, for each selected pair of agents the consensus operation (Definition~\ref{conop})  is applied if and only if the measure of inconsistency between their beliefs, as given in Definition~\ref{inconsistency}, does not exceed a threshold parameter $\gamma \in [0,1]$. Notice that with $\gamma=0$ we have a very conservative model in which only entirely consistent beliefs can be combined, while for the case that $\gamma=1$ we have a model which is equivalent to a totally connected interaction graph, whereby any pair of randomly selected agents may combine their beliefs. In the following, results are presented for a population of $1000$ agents and for the language sizes $|{\cal L}| \in \{1,3,5\}$. The agents' beliefs are initialized by sampling at random from the space of all possible beliefs $\{(x,y)\in [0,1]^2:x \leq y \}^n$. Each run of the simulation is terminated after $50,000$ iterations\footnote{We found that $50,000$ iterations was sufficient to allow simulations to converge across a range of parameter settings.} and the results are averaged over $100$ independent runs.

\begin{figure}[t]
\centering
\includegraphics[width=0.45\textwidth]{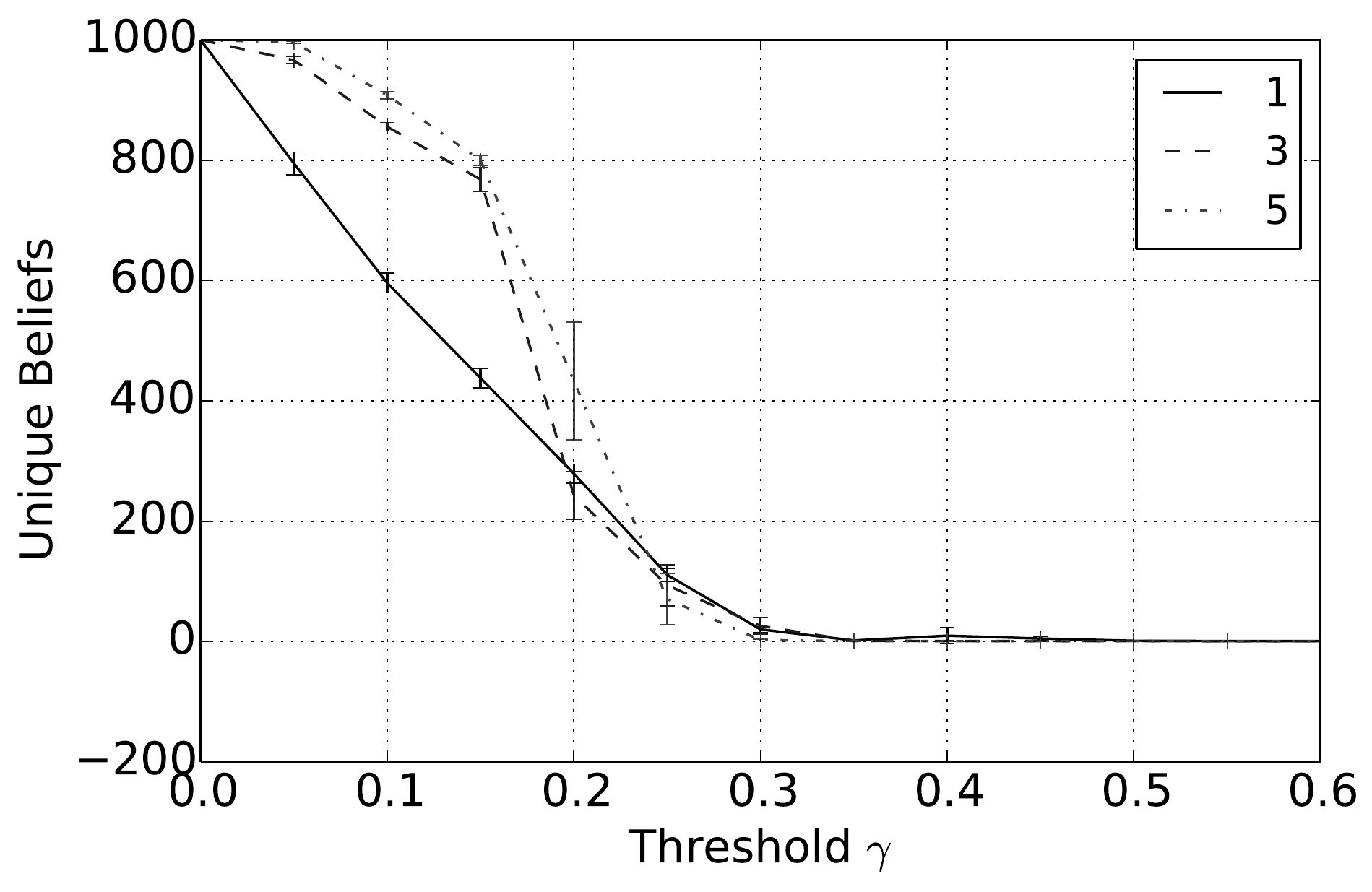}
\caption{Number of unique beliefs after $50,000$ iterations for varying inconsistency thresholds $\gamma$ and various language sizes $|\mathcal{L}|$.}
\label{fig:random_unique_beliefs}
\end{figure}

\begin{figure}[t]
\centering
\includegraphics[width=0.45\textwidth]{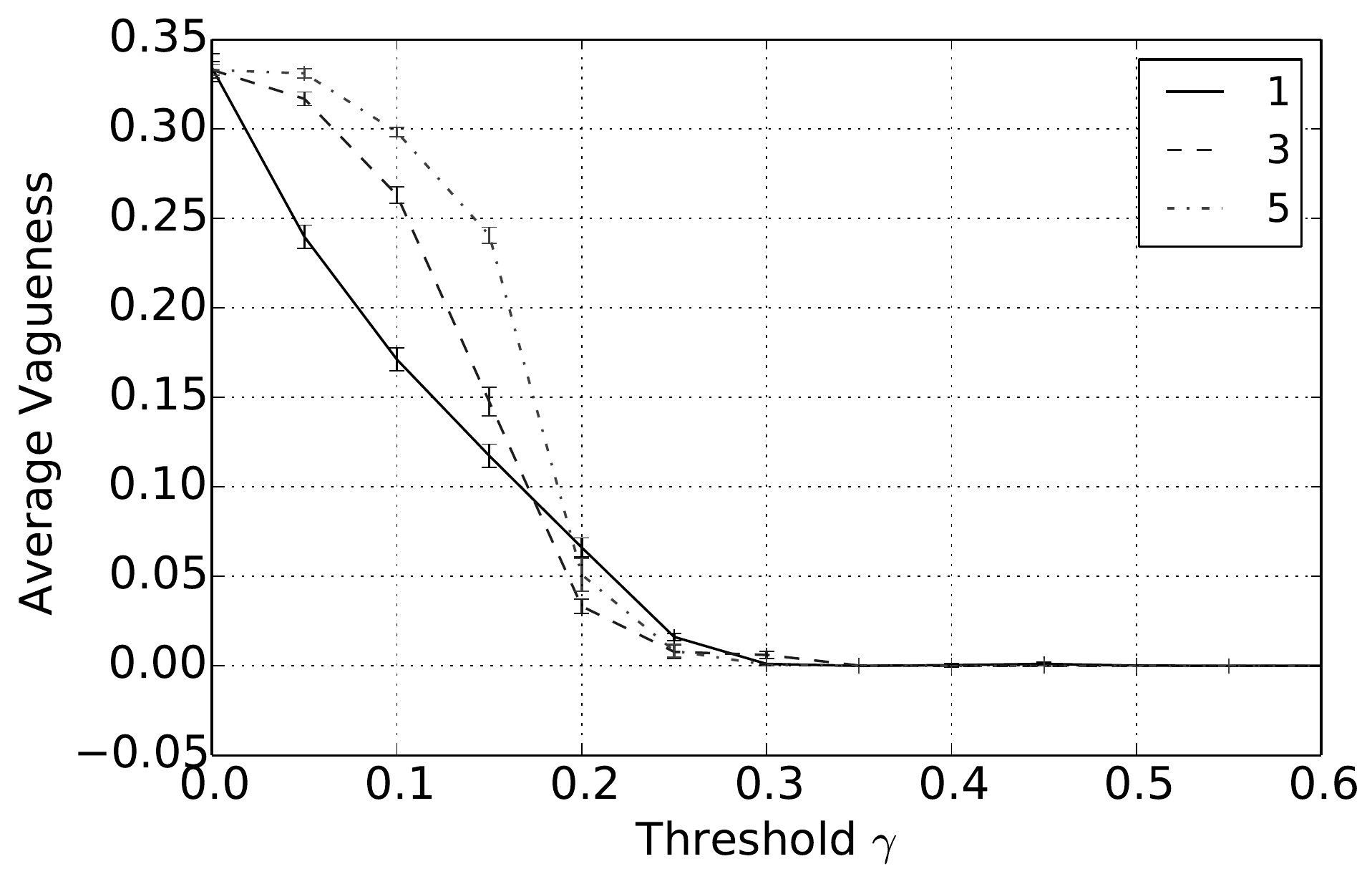}
\caption{Average vagueness after $50,000$ iterations for varying inconsistency thresholds $\gamma$ and various language sizes $|\mathcal{L}|$.}
\label{fig:random_vagueness}
\end{figure}

\begin{figure}[t]
\centering
\includegraphics[width=0.45\textwidth]{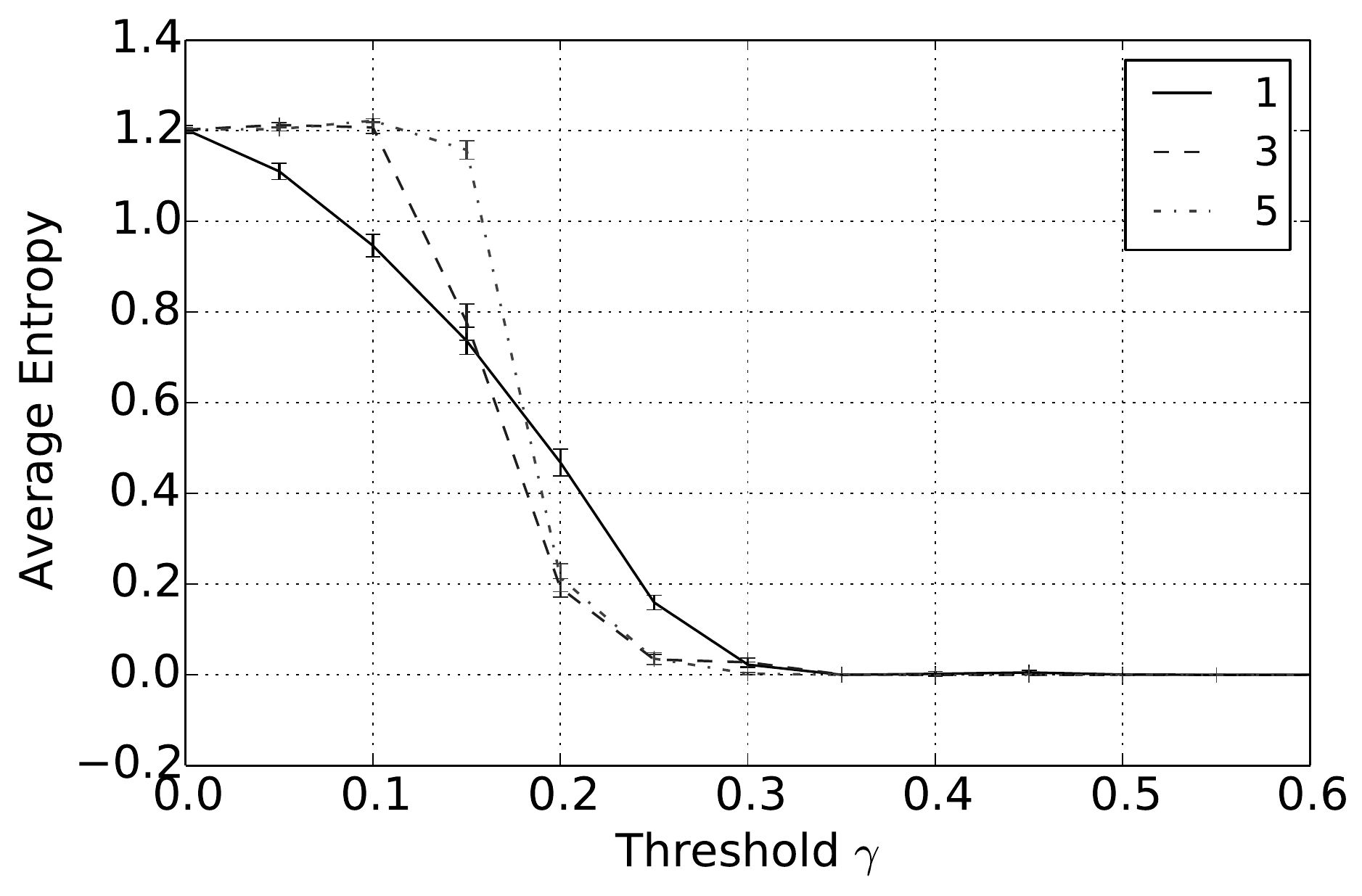}
\caption{Average entropy after $50,000$ iterations for varying inconsistency thresholds $\gamma$ and various language sizes $|\mathcal{L}|$.}
\label{fig:random_entropy}
\end{figure}

Figure~\ref{fig:random_unique_beliefs} shows that the mean number of unique beliefs after $50,000$ iterations decreases with $\gamma$ and for $\gamma \geq 0.5$ there is on average a single belief shared across the population. Furthermore, Figure~\ref{fig:random_vagueness} shows that the vagueness of beliefs, as given in Definition~\ref{vagueness}, averaged both across the different agents and across the independent simulation runs, also decreases with $\gamma$ so that for $\gamma \geq 0.5$ the population has converged to crisp beliefs, i.e. those with a vagueness measure value of $0$. Similarly, from Figure~\ref{fig:random_entropy} we can see that the entropy of beliefs, as given by Definition~\ref{entropy}, decreases with $\gamma$ and for $\gamma \geq 0.5$ at the end of the simulation the population hold beliefs with mean entropy $0$. Hence, summarising Figures~\ref{fig:random_unique_beliefs} to~\ref{fig:random_entropy}, we have that provided the consistency restrictions are sufficiently relaxed, i.e. for $\gamma \geq 0.5$, then a population with beliefs initially allocated at random and with random interactions will converge to a single belief which is both crisp and certain. Unsurprisingly, given the random nature of the agent interactions, the $2^n$ beliefs of this form occur with a uniform distribution across the $100$ independent runs of the simulation.

\begin{figure}[t]
\centering
\includegraphics[width=0.45\textwidth]{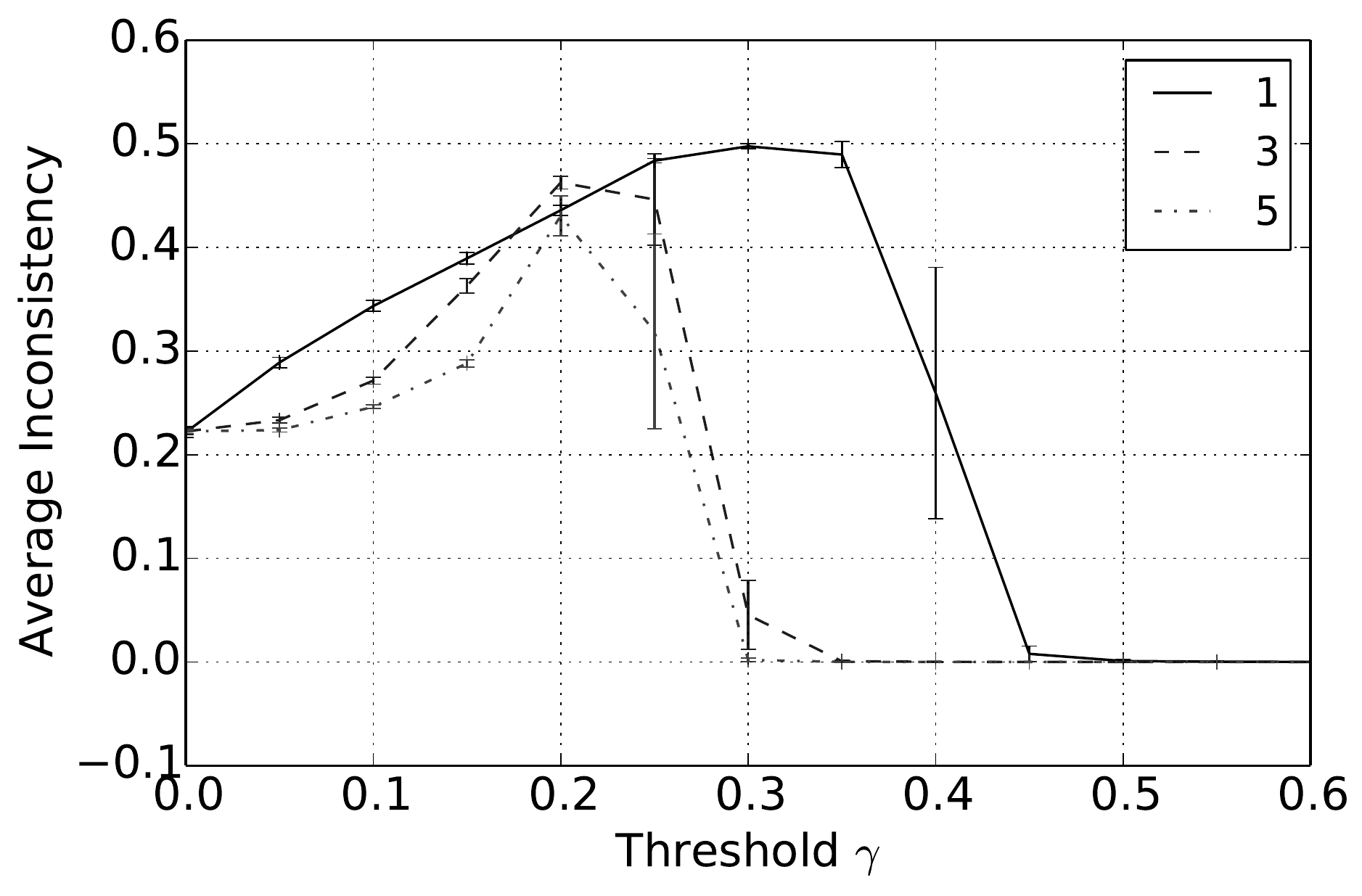}
\caption{Average pairwise inconsistency after $50,000$ iterations for varying inconsistency thresholds $\gamma$ and various language sizes $|\mathcal{L}|$.}
\label{fig:random_inconsistency}
\end{figure}

In addition to the overall consensus reached between agents when $\gamma \geq 0.5$, intermediate values of $\gamma$ between $0.15$ and $0.35$ tend to result in a population with highly polarised opinions. To see this consider figure~\ref{fig:random_inconsistency} showing the average pairwise inconsistency measure value between agents at the end of this simulation and plotted against $\gamma$.  For example, consider the case when $|{\cal L}|=1$ shown as the full black line in Figures~\ref{fig:random_unique_beliefs} to~\ref{fig:random_inconsistency}. In this case we see that the mean inconsistency value obtains a maximum of $0.5$ at around $\gamma=0.28$. Furthermore, from Figures~\ref{fig:random_unique_beliefs} to~\ref{fig:random_entropy} we see that for this value of $\gamma$ the average number of unique beliefs, the vagueness, and entropy are all relatively low. Consequently, we are seeing a polarisation of opinions where individuals are holding a small number of highly inconsistent beliefs which are also relatively crisp and certain. Such behaviour, while still present, is less pronounced for language sizes $|{\cal L}|=3$ and $5$. This may be due to the fact that, since Definition~\ref{inconsistency} is an average of inconsistency values across the propositions in ${\cal L}$, increasing the language size reduces the variance of the inconsistency values in the initial population. Furthermore, as $|{\cal L}|$ increases the distribution of inconsistency values is approximately normal with mean $\frac{2}{9}$.
Hence, for $\gamma \geq \frac{2}{9}$ the probability that a randomly selected pair of agents will have an inconsistency value exceeding $\gamma$ decreases as $|{\cal L}|$ increases. This in turn will increase the probability of agreement in any interaction, reducing the likelihood of opinion polarisation for $\gamma \geq \frac{2}{9}$.

\section{Simulation Experiments Involving Consensus Formation and Belief Updating}
In~\cite{hegselmann05} an opinion model is investigated in which agents receive direct evidence about the state of the world, perhaps from an ongoing measurement process, as well as pooling the opinion of others with similar beliefs. The original model in~\cite{hegselmann05} involves real valued beliefs but this has been adapted by~\cite{reigler} to the case in which beliefs and evidence are theories in a propositional logic language. The fundamental question under consideration is whether or to what extent dialogue between individuals, for example scientists, helps them to find the truth or instead whether they are better off simply to wait until they receive direct evidence? In this section we investigate this question in the context of vague and uncertain beliefs and where consensus building is modelled using the combination operator in Definition~\ref{conop}. Direct evidence is then provided to the population at random instances when an individual is told the truth value of a proposition. That agent then updates her beliefs by adopting a compromise position between her current opinions and the evidence provided.

\begin{figure}[t]
\centering
\includegraphics[width=0.45\textwidth]{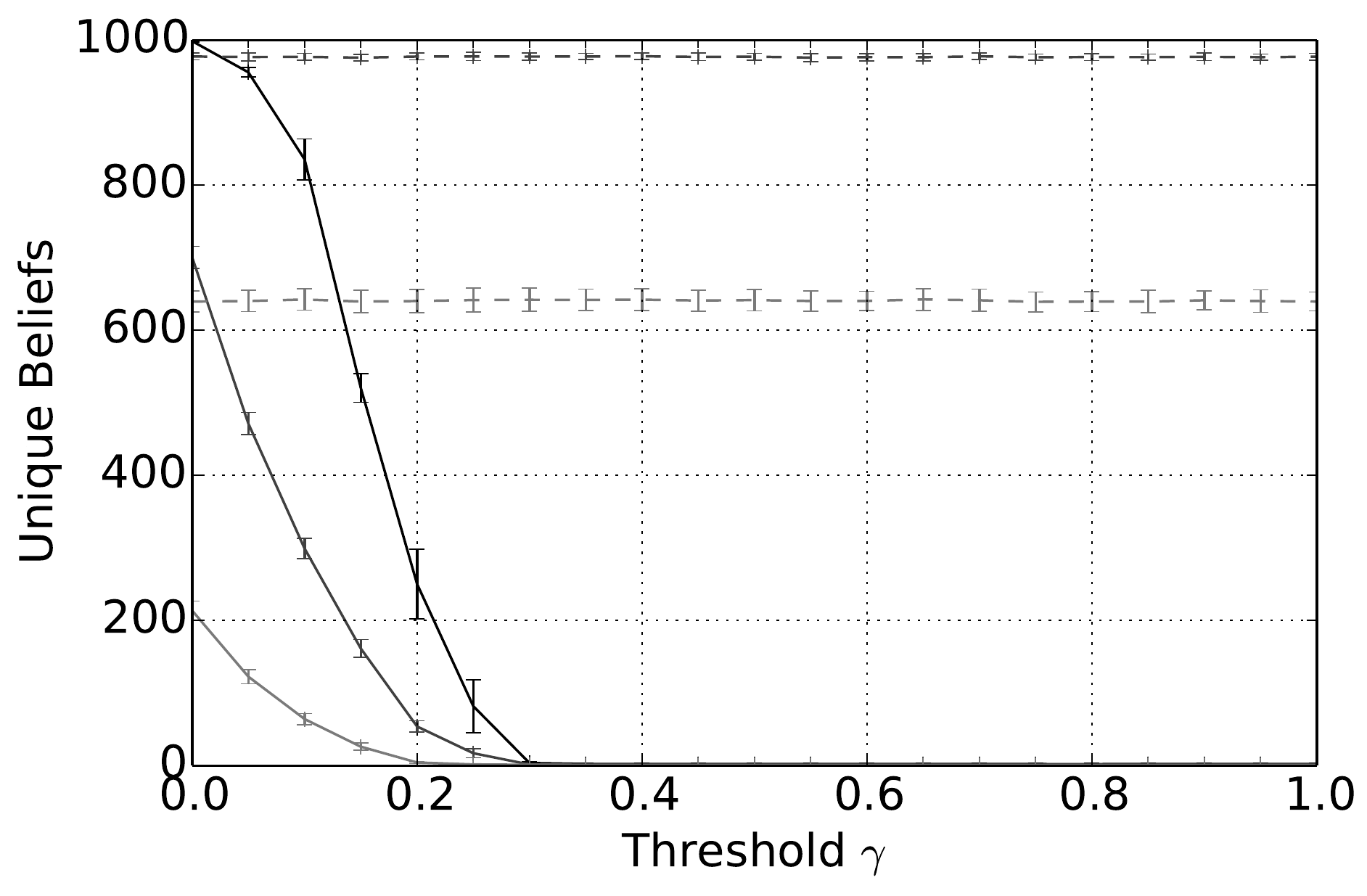}
\put(-197,110){\footnotesize Both ($5 \%$)}
\put(-145,130){\footnotesize Evidence only ($5\%$, $15\%$)}
\put(-192,75){\footnotesize \shortstack[l]{Both \\($15 \%$)} }
\put(-198,40){\footnotesize \shortstack[l]{Both \\($30 \%$)} }
\put(-145,105){\footnotesize Evidence only ($30\%$)}
\caption{Number of unique beliefs after $50,000$ iterations for varying inconsistency thresholds $\gamma$, $|\mathcal{L}| = 5$ and evidence rates $\alpha = 5,15$ and $30\%$. The solid lines refer to evidential updating combined with consensus building while the dotted lines refer to evidential updating only.}
\label{fig:unique_beliefs_evidence}
\end{figure}
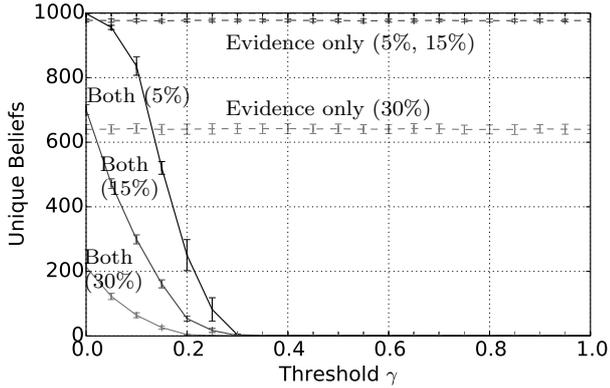

We assume that the true state of the world is a Boolean valuation $\mathbf{v}^*$ on ${\cal L}$ so that $\mathbf{v}^*(p_i) \in \{0,1\}$ for $i=1, \ldots, n$. Now given the discussion in section 2 about interpreting the third truth value as meaning `borderline', this is clearly a simplification from that perspective. For example, consider the proposition `Ethel is short', then an experiment could consist of measuring Ethel's height according to some mechanism, and then comparing it to the experimenter's definition of the term `short' in order to determine the truth value of the proposition. If that definition is three-valued then the outcome of the experiment could well be to identify a borderline truth value for the proposition. However, the convention in science is to establish an agreed crisp definition of all the terms used to express a hypothesis so that the resulting proposition is falsifiable. This would then be consistent with our identifying the true state of the world with a Boolean valuation.  In the following definition we propose a measure of belief quality which quantifies the similarity of an agent's beliefs to the true state of the world. This will subsequently be used to assess the extent to which the population has converged to the truth. Furthermore, it will also be employed in the next section as a mechanism for providing indirect information about the state of the world.

\begin{definition}{A Quality Measure}\\
\label{payoff}
Let $f:{\cal L} \rightarrow \{-1,1\}$ be such that $f(p_i)=2\mathbf{v}^*(p_i)-1$ is the payoff for believing that $p_i$ has truth value $1$ and $-f(p_i)$ is the payoff for believing that the truth value of $p_i$ is $0$. Furthermore, it is always assumed that believing that $p_i$ has truth value $\frac{1}{2}$ has payoff $0$. Then we define the quality or payoff for the belief $\boldsymbol{\mu}$ by:
\begin{gather*}
\sum_{i=1}^n \left( f(p_i)(\underline{\mu}(p_i)+\overline{\mu}(p_i)-1 ) \right)
\end{gather*}
\end{definition}
Notice that $f(p_i)(\underline{\mu}(p_i)+\overline{\mu}(p_i)-1 )=f(p_i)\underline{\mu}(p_i)+(- f(p_i))(1-\overline{\mu}(p_i))$ corresponding to the agent's expected payoff from their beliefs about proposition $p_i$. Definition~\ref{payoff} then takes the sum of this expected payoff across the propositions in ${\cal L}$.

\begin{figure}[t]
\centering
\includegraphics[width=0.45\textwidth]{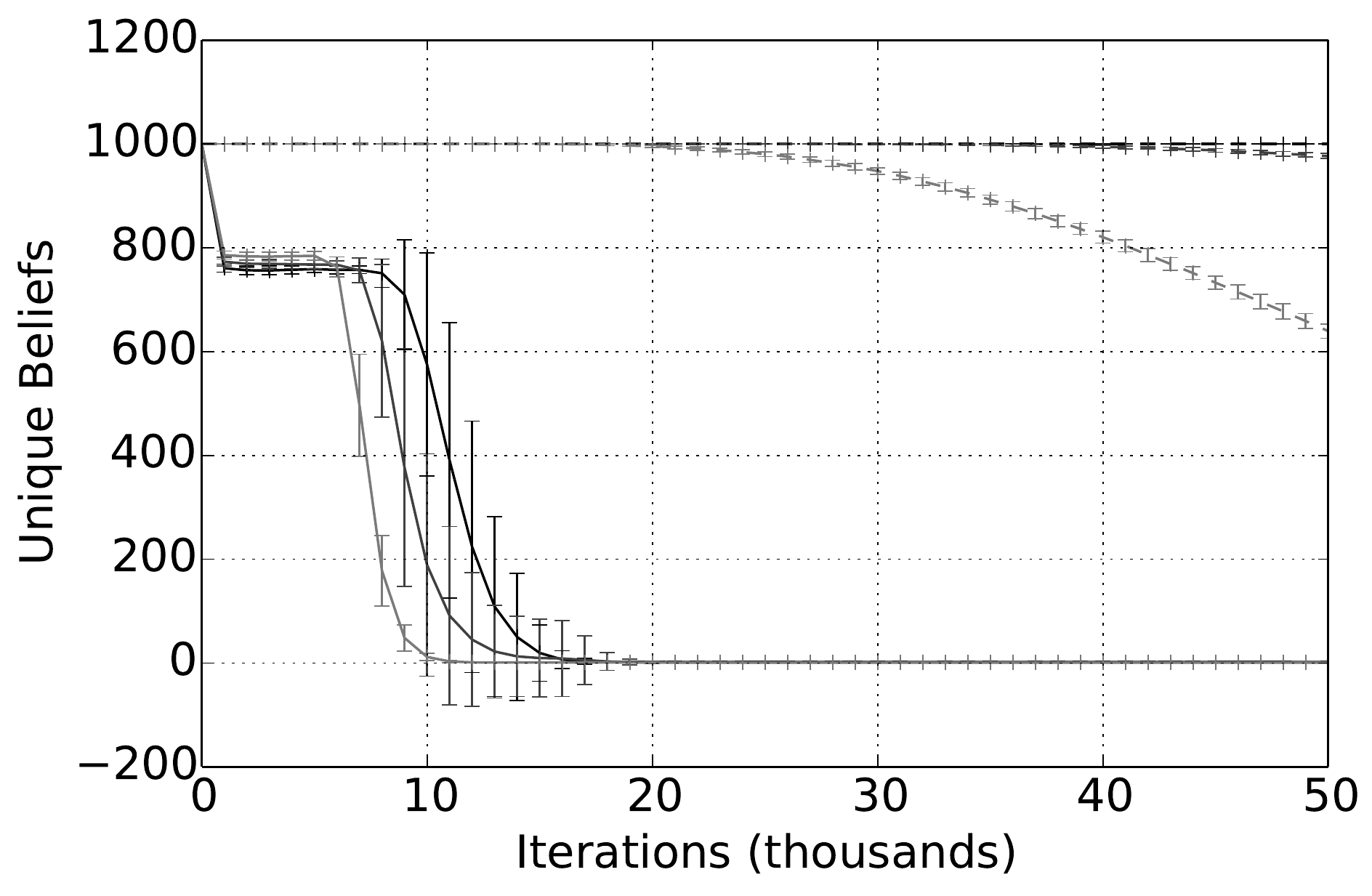}
\caption{Number of unique beliefs over $50,000$ iterations for $\gamma = 0.8$, $|\mathcal{L}| = 5$ and evidence rates $\alpha = 5,15$ and $30\%$. The solid lines refer to evidential updating combined with consensus building while the dotted lines refer to evidential updating only.}
\label{fig:unique_beliefs_evidence_trajectory}
\end{figure}

The simulations consist of $1000$ agents with beliefs initially picked at random from $\{(x,y)\in [0,1]^2: x \leq y \}^n$ as in section 4. Furthermore, the true state of the world $\mathbf{v}^*$ is picked at random from $\{0,1\}^n$ prior to the simulation and the payoff $f$ calculated as in Definition~\ref{payoff}. Each run of the simulation is terminated after $50,000$ iterations and the results are averaged over $100$ runs. At each iteration two agents are selected at random and apply the consensus operator (Definition~\ref{conop}) provided that the inconsistency level of their current beliefs does not exceed $\gamma$. Furthermore, at each iteration there is a fixed $\alpha \%$ chance of direct evidence being presented to the population. In the case that it is, an agent is selected at random and told the value of $\mathbf{v}^*(p_i)$ for some proposition also selected at random from those in ${\cal L}$. The agent then updates her current beliefs $\boldsymbol{\mu}$ to $\boldsymbol{\mu}^\prime$ where
\begin{gather*}
\boldsymbol{\mu}^\prime =\boldsymbol{\mu} \odot \left( (0,1), \ldots, (\mathbf{v}^*(p_i),\mathbf{v}^*(p_i)), \ldots, (0,1) \right)
\end{gather*}
In other words, the agent adopts a new set of beliefs formed as a compromise between her current beliefs and the evidence, the latter being interpreted as a set of beliefs where $\boldsymbol{\mu}(p_i)=(\mathbf{v}^*(p_i),\mathbf{v}^*(p_i))$ and $\boldsymbol{\mu}(p_j)=(0,1)$ for $j \neq i$. That is they form consensus with an alternative opinion which is certain about the truth value of $p_i$ and is  neutral about the other propositions. Notice that in this case it follows from Definition~\ref{conop} that $\boldsymbol{\mu}^\prime(p_i)=(\overline{\mu}(p_1),1)$ if $\mathbf{v}^*(p_i)=1$, $\boldsymbol{\mu}^\prime(p_i)=(0,\underline{\mu}(p_i))$ if $\mathbf{v}^*(p_i)=0$ and $\boldsymbol{\mu}^\prime(p_j)=\boldsymbol{\mu}(p_j)$ for $j \neq i$. The combined consensus and evidential belief updating approach can then be compared with simulations in which only the above belief updating model is applied and in which there is no consensus building.

\begin{figure}[t]
\centering
\includegraphics[width=0.45\textwidth]{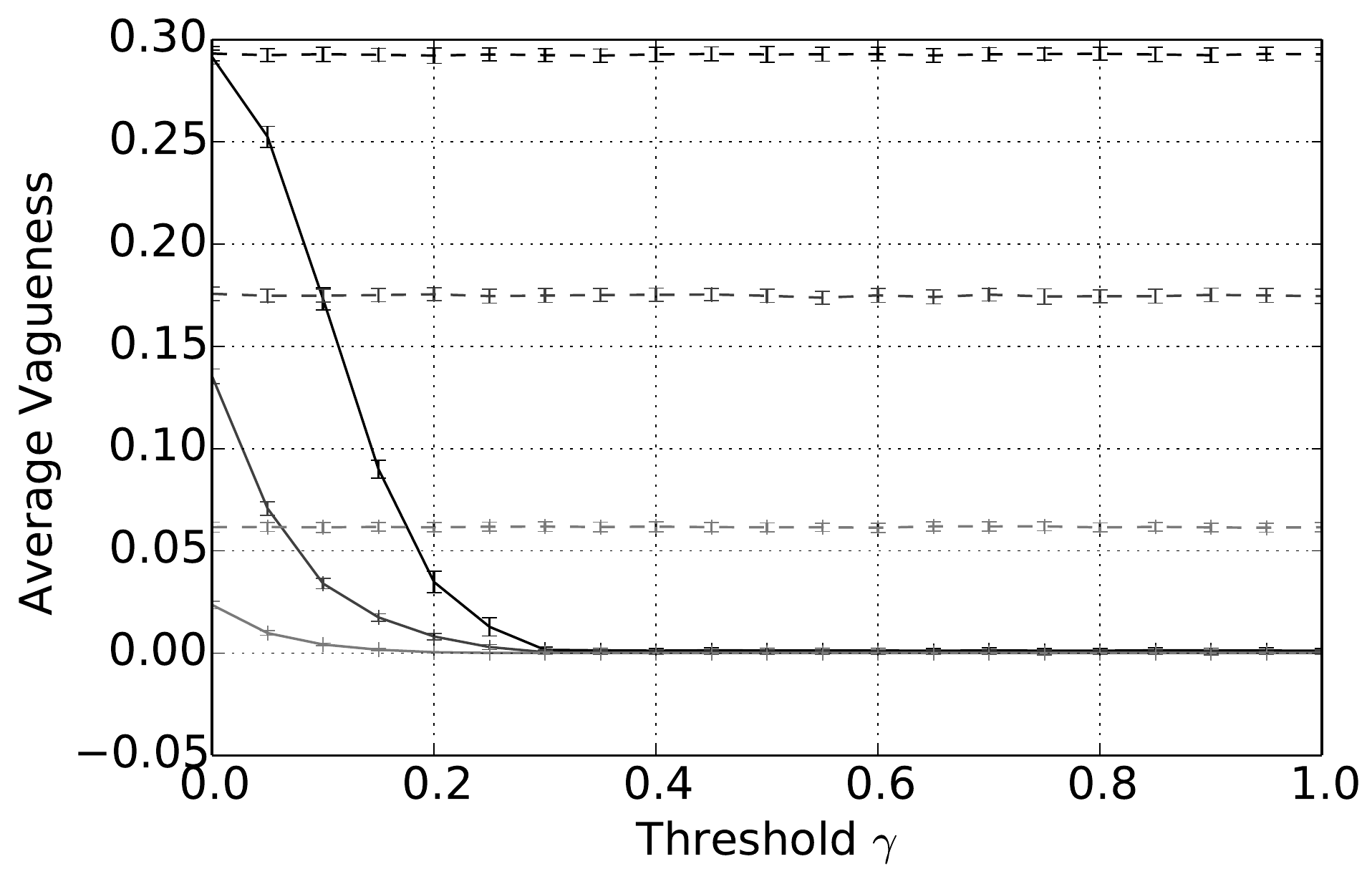}
\put(-195,108){\footnotesize Both ($5 \%$)}
\put(-145,130){\footnotesize Evidence only ($5\%$)}
\put(-195,75){\footnotesize \shortstack[l]{Both \\($15 \%$)} }
\put(-145,105){\footnotesize Evidence only ($15\%$)}
\put(-195,43){\footnotesize \shortstack[l]{Both \\($30 \%$)} }
\put(-145,62){\footnotesize Evidence only ($30\%$)}
\caption{Average vagueness after $50,000$ iterations for varying inconsistency thresholds $\gamma$, $|\mathcal{L}| = 5$ and evidence rates $\alpha = 5,15$ and $30\%$. The solid lines refer to evidential updating combined with consensus building while the dotted lines refer to evidential updating only.}
\label{fig:vagueness_evidence}
\end{figure}
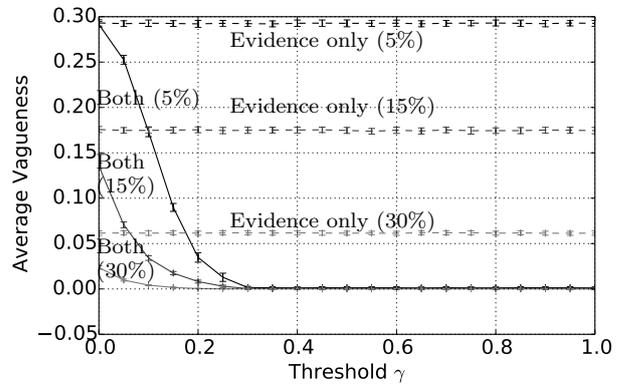

In this section we focus on evidence rates of $\alpha=5, \ 15$ and $30\%$ and we assume that the language size is $|{\cal L}|=5$. For these parameter settings Figure~\ref{fig:unique_beliefs_evidence} shows that for $\gamma \geq 0.4$, all three cases in which evidential updating is combined with consensus formation converge on shared belief across the population. Furthermore, the higher the evidence rate $\alpha$, the greater the convergence for any given threshold value $\gamma$. It is also clear from Figure~\ref{fig:unique_beliefs_evidence} that combining consensus building with evidential updating leads to much better convergence than evidence based updating alone. For instance, we see that for evidential updating alone it is only with an evidence rate of $30\%$ that there is a large reduction in the number of distinct beliefs in the population after $50,000$ iterations, with the population still containing over $900$ different opinions for both the $5\%$ and $15\%$ rates. Furthermore, Figure~\ref{fig:unique_beliefs_evidence_trajectory} shows a typical trajectory for the average number of unique beliefs against iteration when $\gamma=0.8$. Notice that after $25,000$ iterations all three of the combined models have converged to a single shared belief. In contrast the evidence-only approaches have still not converged after $50,000$ iterations, where even with a $30\%$ evidence rate there are still over $600$ distinct opinions remaining in the population.

\begin{figure}[t]
\centering
\includegraphics[width=0.45\textwidth]{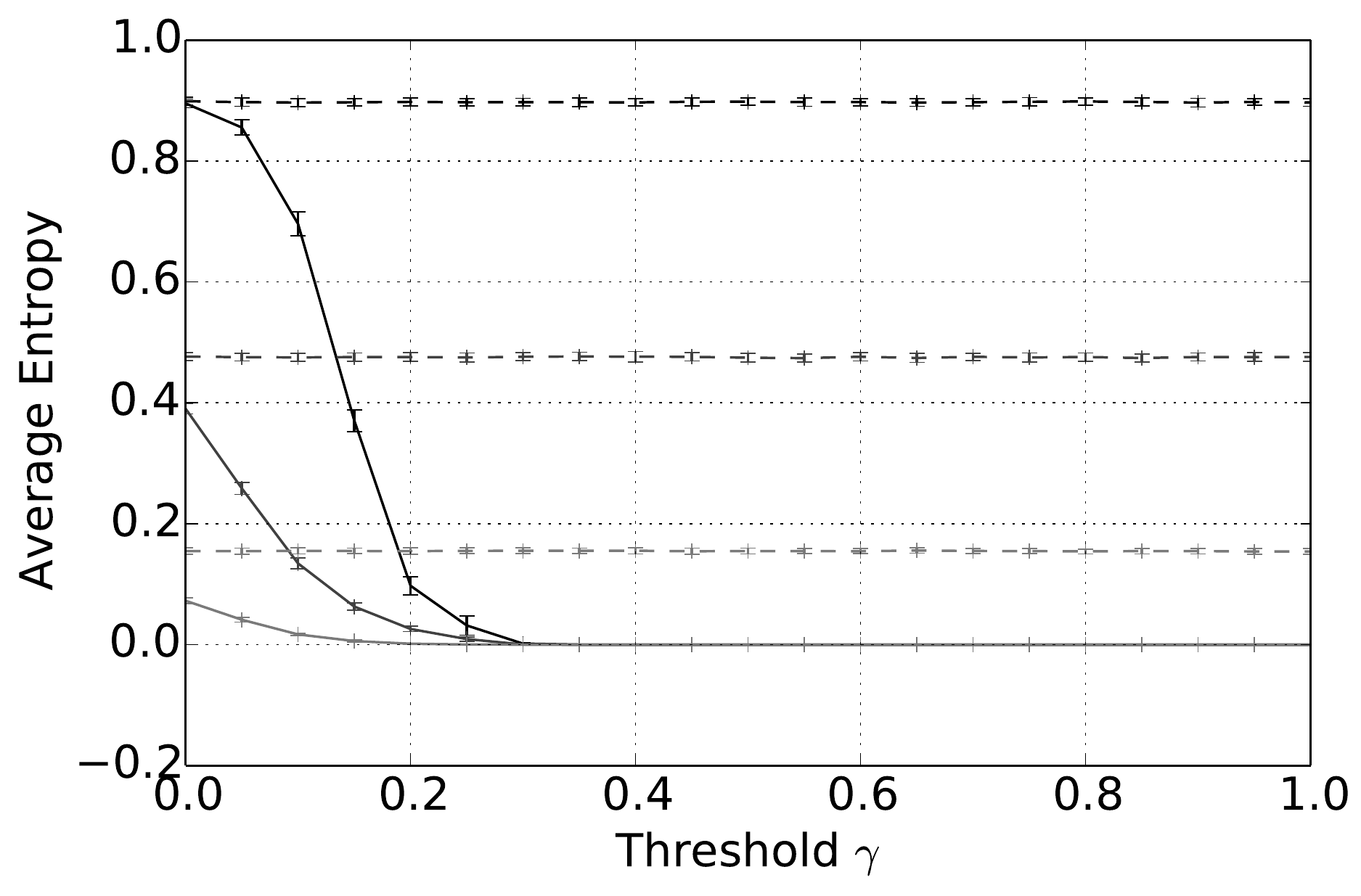}
\put(-189,125){\footnotesize Both ($5 \%$)}
\put(-145,137){\footnotesize Evidence only ($5\%$)}
\put(-195,70){\footnotesize \shortstack[l]{Both \\($15 \%$)} }
\put(-145,95){\footnotesize Evidence only ($15\%$)}
\put(-195,40){\footnotesize \shortstack[l]{Both \\($30 \%$)} }
\put(-145,62){\footnotesize Evidence only ($30\%$)}
\caption{Average entropy after $50,000$ iterations for varying inconsistency thresholds $\gamma$, $|\mathcal{L}| = 5$ and evidence rates $\alpha = 5,15$ and $30\%$. The solid lines refer to evidential updating combined with consensus building while the dotted lines refer to evidential updating only.}
\label{fig:entropy_evidence}
\end{figure}
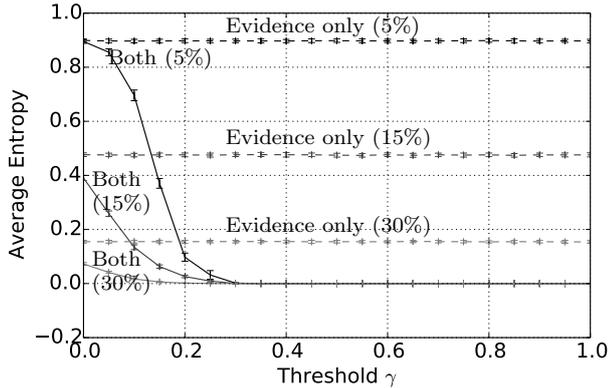

\begin{figure}[t]
\centering
\includegraphics[width=0.45\textwidth]{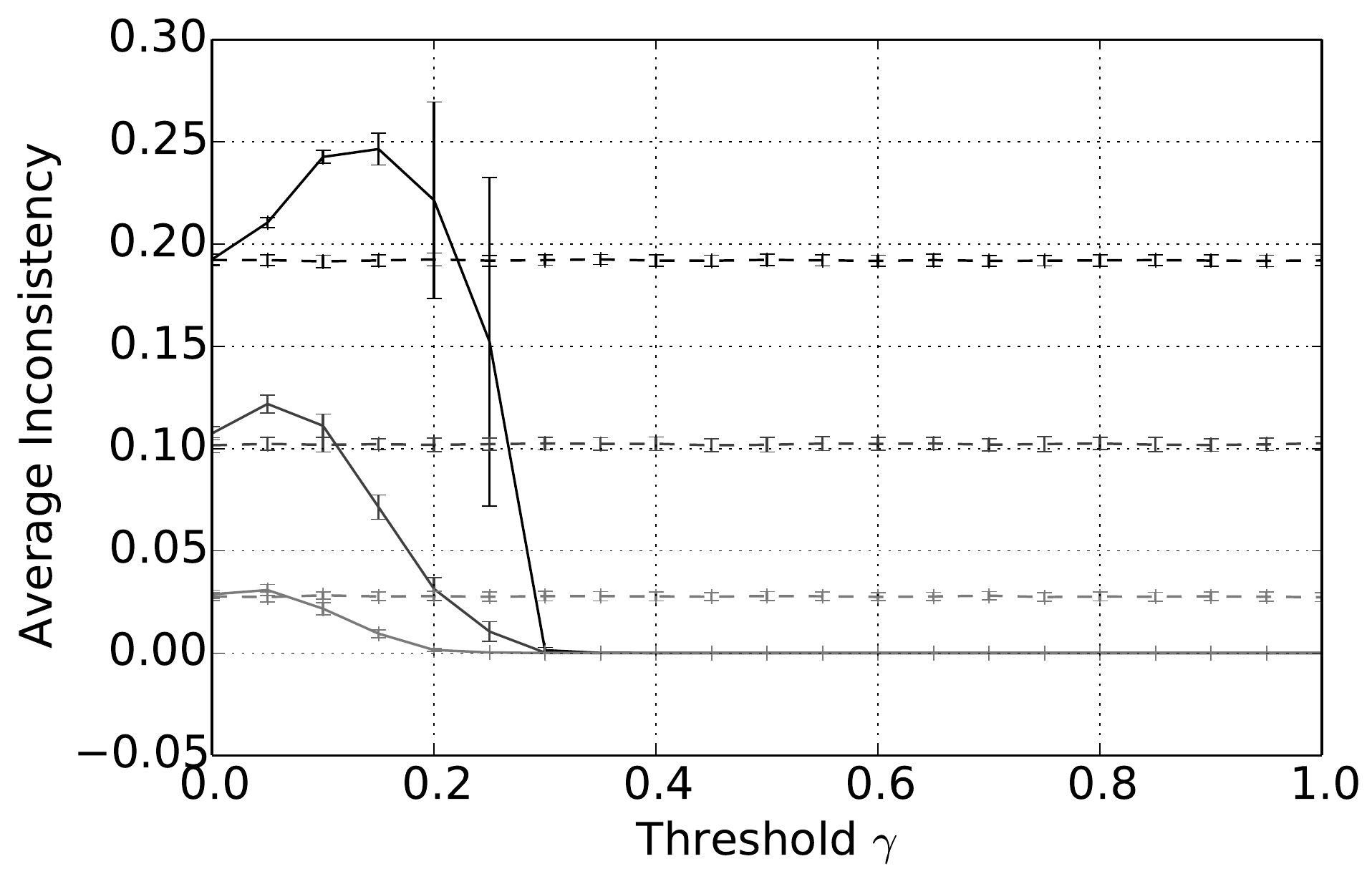}
\put(-195,130){\footnotesize Both ($5 \%$)}
\put(-135,110){\footnotesize Evidence only ($5\%$)}
\put(-195,82){\footnotesize \shortstack[l]{Both \\($15 \%$)} }
\put(-135,77){\footnotesize Evidence only ($15\%$)}
\put(-195,50){\footnotesize \shortstack[l]{Both \\($30 \%$)} }
\put(-135,50){\footnotesize Evidence only ($30\%$)}
\caption{Average pairwise inconsistency after $50,000$ iterations for varying inconsistency thresholds $\gamma$, $|\mathcal{L}|=5$ and evidence rates $\alpha = 5,15$ and $30\%$. The solid lines refer to evidential updating combined with consensus building while the dotted lines refer to evidential updating only.}
\label{fig:inconsistency_evidence}
\end{figure}

\begin{figure}[t]
\centering
\includegraphics[width=0.45\textwidth]{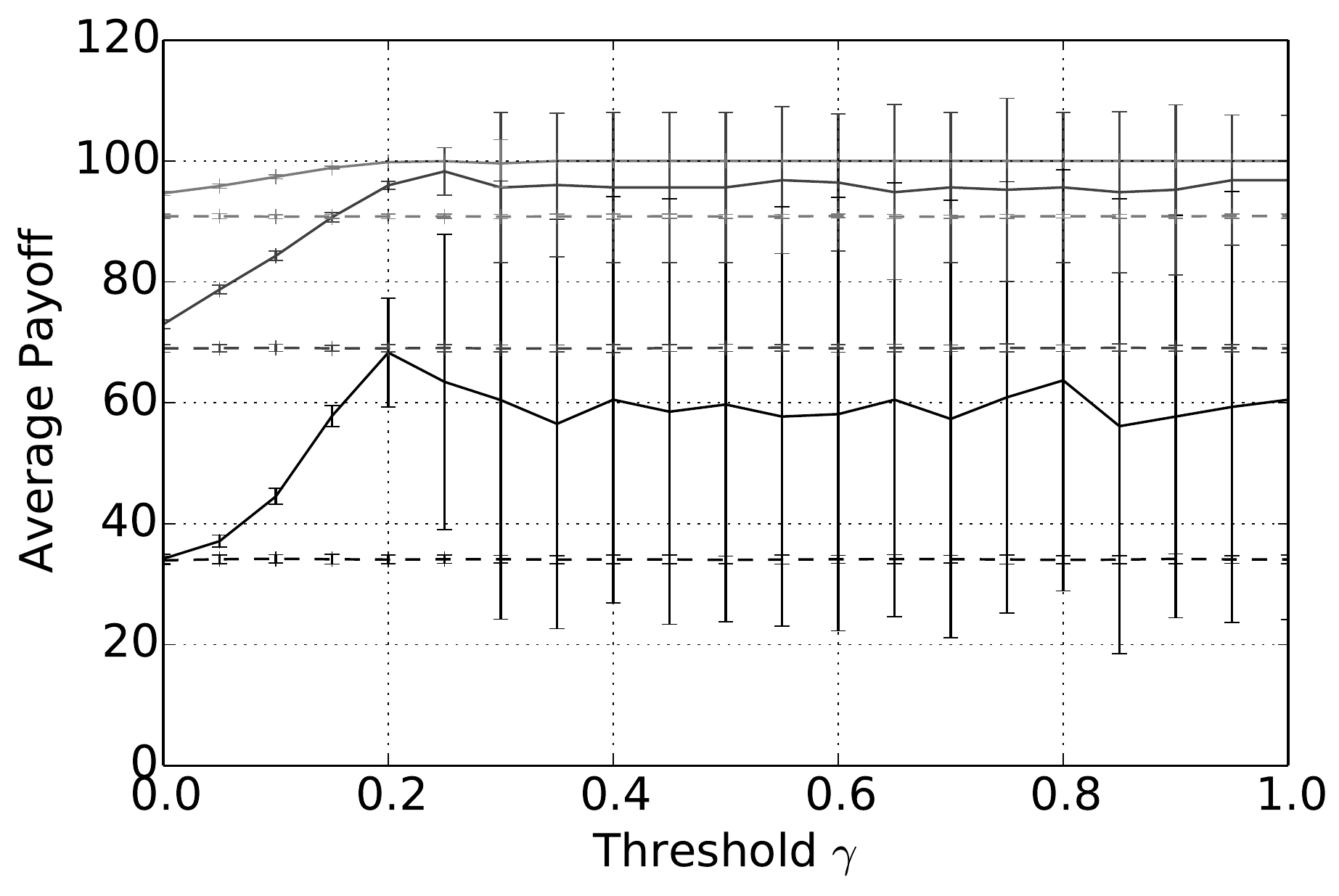}
\put(-202,130){\footnotesize Both ($30\%$)}
\put(-130,108){\footnotesize Evidence only ($30 \%$)}
\put(-202,98){\footnotesize \shortstack[l]{Both \\($15 \%$)} }
\put(-130,96){\footnotesize Evidence only ($15\%$)}
\put(-202,67){\footnotesize \shortstack[l]{Both \\($5 \%$)} }
\put(-130,60){\footnotesize Evidence only ($5\%$)}
\caption{Average payoff after $50,000$ iterations for varying inconsistency thresholds $\gamma$, $|\mathcal{L}|=5$ and evidence rates $\alpha = 5,15$ and $30\%$. The solid lines refer to evidential updating combined with consensus building while the dotted lines refer to evidential updating only.}
\label{fig:payoff_evidence}
\end{figure}

Taken together with Figure~\ref{fig:unique_beliefs_evidence}, Figures~\ref{fig:vagueness_evidence} and~\ref{fig:entropy_evidence} show that, assuming a sufficiently high threshold value $\gamma \geq 0.4$, the combined consensus building and updating approach results in convergence to a shared belief which is both crisp and certain. Again, increasing the evidence rate leads to a reduction in both the average vagueness and average entropy for any given threshold value and evidence based updating alone results in much higher values for the same evidence rate. The overall convergence of the population is also shown by the average pairwise inconsistency values in Figure~\ref{fig:inconsistency_evidence}. The convergence of the combined approach to a shared opinion for all evidence rates and thresholds $\gamma \geq 0.4$ is reflected in a zero average inconsistency level for this range of parameters.  Notice, however, that for all evidence rates the average inconsistency for the combined approach has a peak value in the range $0 < \gamma < 0.4$, suggesting that there is some polarisation of opinion for thresholds in this range. For evidence updating only the level of inconsistency is relatively higher than for the combined approach for all evidence rates suggesting that there is a much higher level of disagreement remaining between agents after $50,000$ iterations. Finally, Figure~\ref{fig:payoff_evidence} shows the average payoff values calculated as in Definition~\ref{payoff} and given as a percentage of the maximum possible value i.e. in this case $5$. These values reflect the extent to which the population have converged to a set of beliefs close to the true state of the world. For each of the three evidence rates, given a sufficiently high threshold value, the combined approach results in an average payoff which is significantly higher than for evidential updating alone. Indeed for a $30\%$ evidence rate and $\gamma \geq 0.3$ the combination of consensus building and belief updating results in close to the maximum payoff value on average i.e. the population has learnt the state of the world with an average accuracy of close to $100\%$.

\section{Simulation Experiments with Agent Selection Influenced by Belief Quality.}

\begin{figure}[t]
\centering
\includegraphics[width=0.45\textwidth]{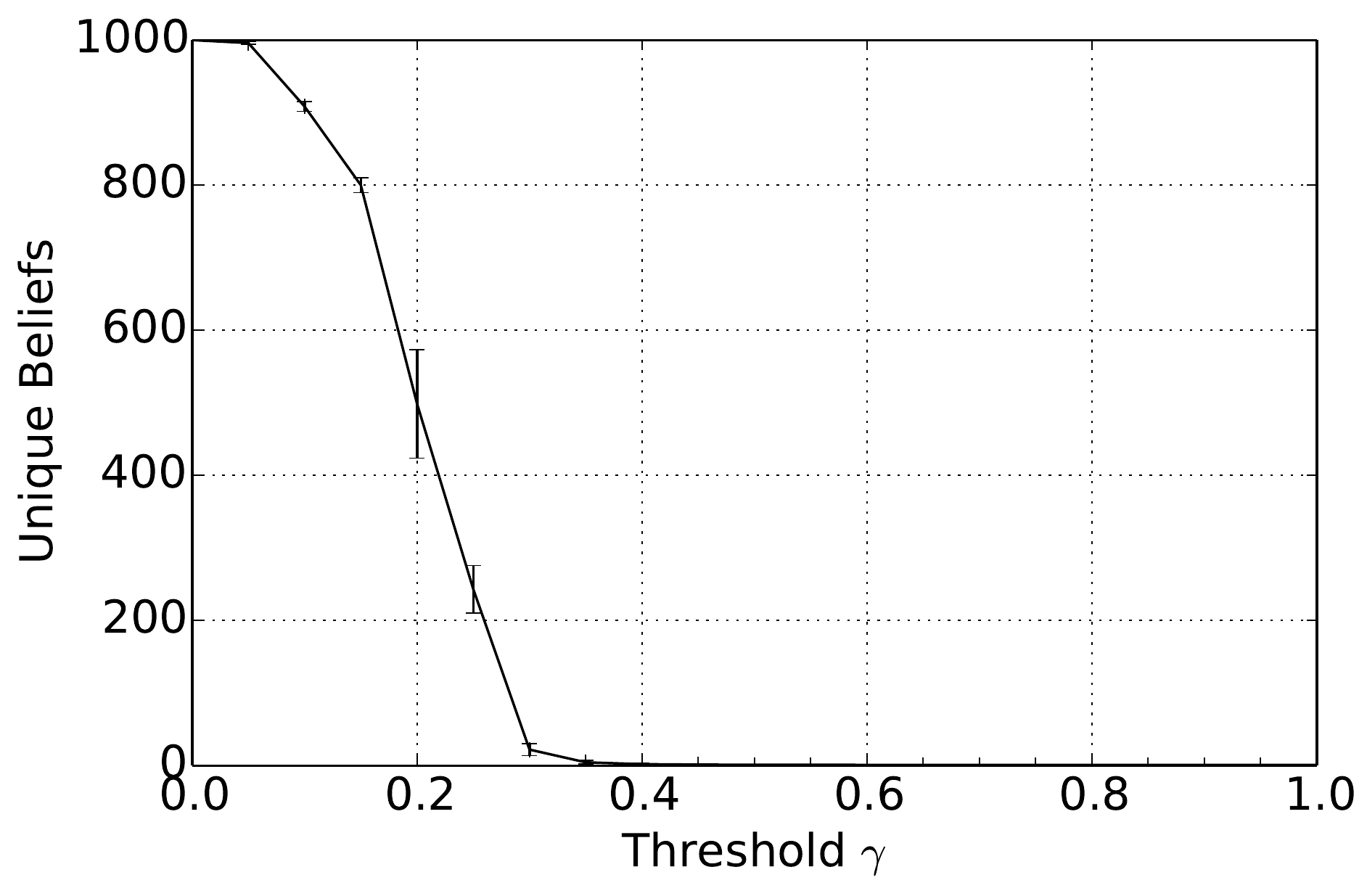}
\caption{Number of unique beliefs after $50,000$ iterations for varying inconsistency thresholds $\gamma$ and $|\mathcal{L}| = 5$.}
\label{fig:payoff_unique_beliefs}
\end{figure}



In this section we consider a scenario in which agents receive indirect feedback about the accuracy of their beliefs in the form of payoff or reward obtained as a result of actions that they have taken on the basis of these beliefs. Furthermore, we assume that the closer that an agent's beliefs are to the actual state of the world then the higher their rewards will be on average. Hence, we use the payoff measure given in Definition~\ref{payoff} as a proxy for this process so that agent selection in the consensus building process is guided by the payoff or quality measure of their beliefs. More specifically, we now investigate an agent-based system in which pairs of agents are selected for interaction with a probability that is proportional to the product of the quality of their respective beliefs. For modelling societal opinion dynamics this captures an assumption that better performing agents, i.e. those with higher payoff, are more likely to interact in a context in which both parties will benefit from reaching an agreement. In biological systems there are examples of a similar quality effect on distributed decision making. For instance, honeybee swarms collectively choose between alternative nesting sites by means of a dance in which individual bees indicate the direction of the site that they have just visited~\cite{list}. The duration of the dance is dependant on the quality of the site and this in turn affects the likelihood that the dancer will influence other bees. Artificial systems can of course be designed so that interactions are guided by quality provided that a suitable measure of the latter can de defined, as is typically the case in evolutionary computing.

\begin{figure}[t]
\centering
\includegraphics[width=0.45\textwidth]{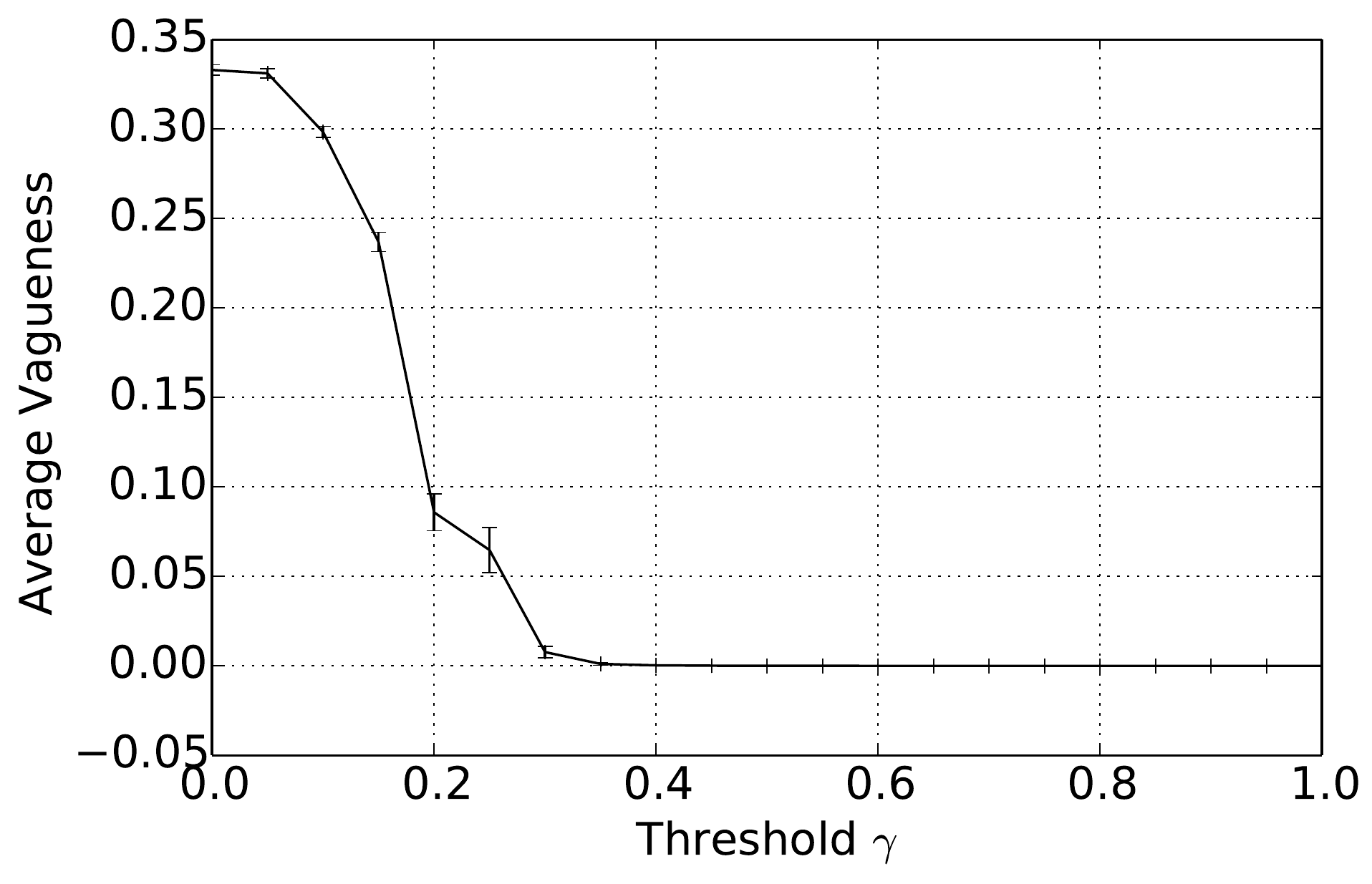}
\caption{Average vagueness after $50,000$ iterations for varying inconsistency thresholds $\gamma$ and $|\mathcal{L}| = 5$.}
\label{fig:payoff_vagueness}
\end{figure}



We now describe the results from running agent-based simulations mainly following the same template as described in section 4 but with an important difference. Instead of being selected at random, agents were instead selected for interaction with probability proportional to the quality value of their beliefs as given in Definition~\ref{payoff}. The true state of the world $\mathbf{v}^*$ was chosen at random from $\{0,1\}^n$ prior to running the simulation and the payoff function $f$ was then determined as in Definition~\ref{payoff}. As in the previous sections the population consisted of $1000$ agents with initial beliefs selected at random from $\{(x,y) \in [0,1]:x \leq y \}^n$. All result in this section relate to the language size $|{\cal L}|=5$.

\begin{figure}[t]
\centering
\includegraphics[width=0.45\textwidth]{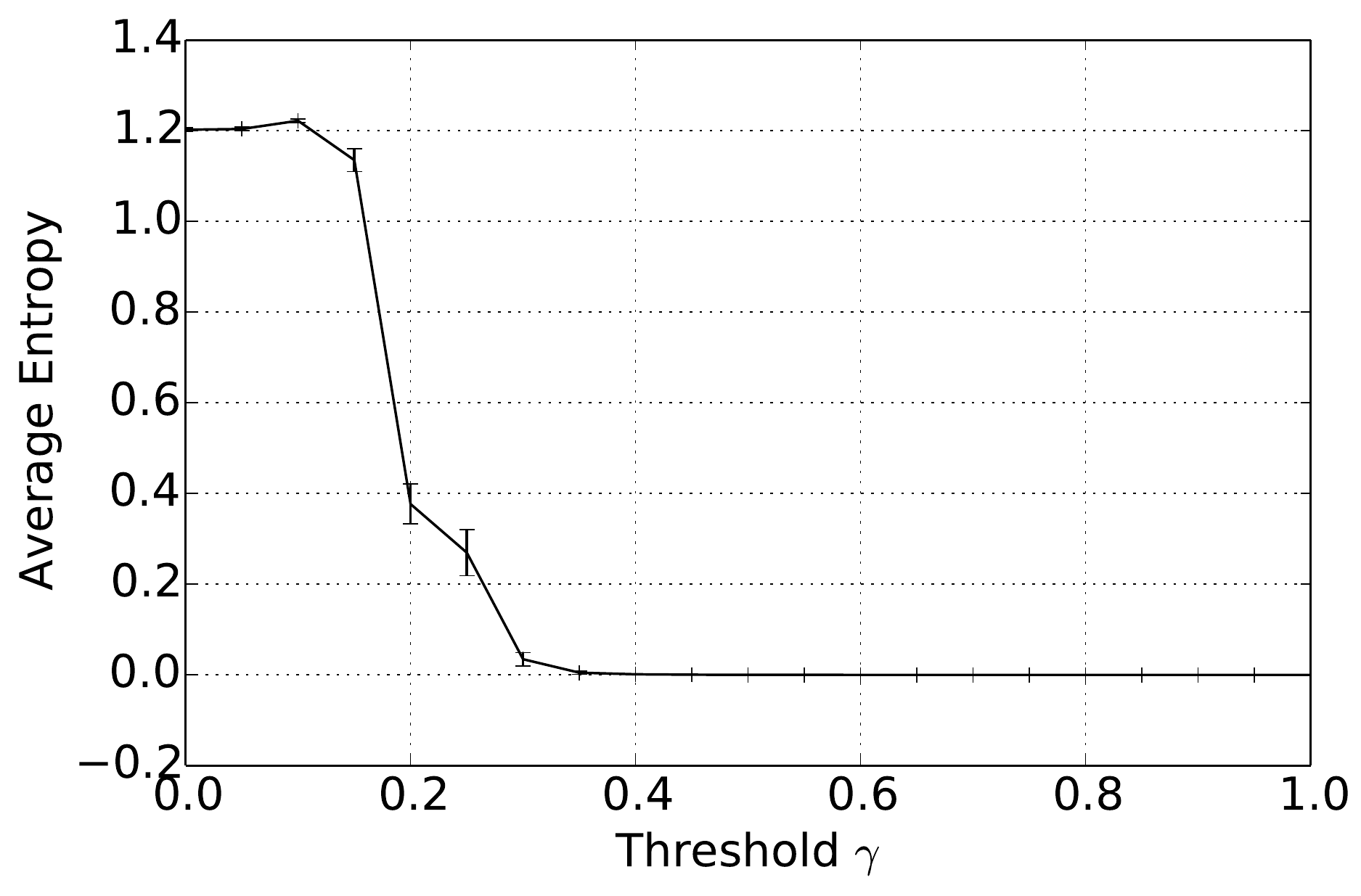}
\caption{Average entropy after $50,000$ iterations for varying inconsistency thresholds $\gamma$ and $|\mathcal{L}| = 5$.}
\label{fig:payoff_entropy}
\end{figure}

\begin{figure}[t]
\centering
\includegraphics[width=0.45\textwidth]{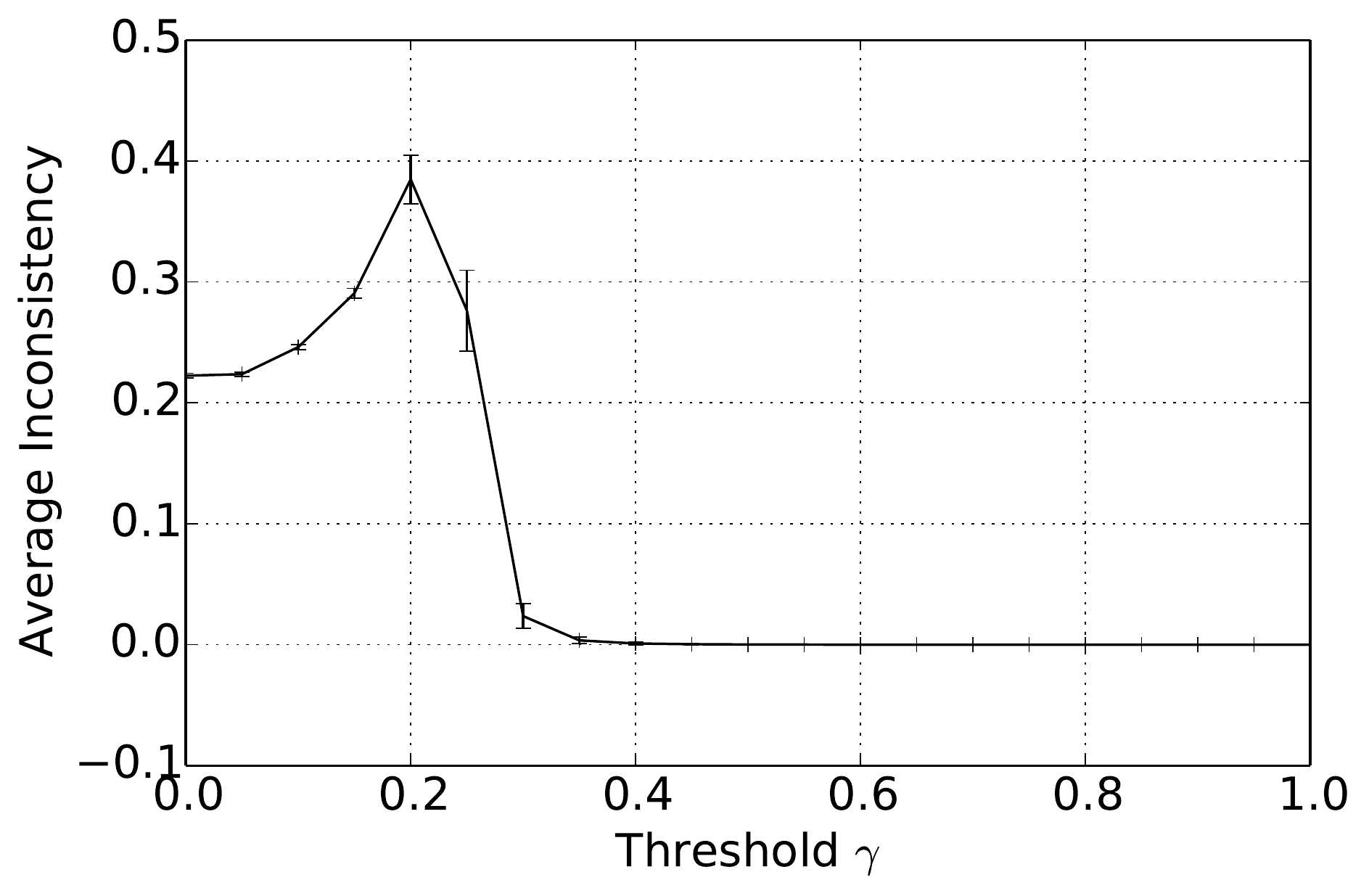}
\caption{Average pairwise inconsistency after $50,000$ iterations for varying inconsistency thresholds $\gamma$ and $|\mathcal{L}| = 5$.}
\label{fig:payoff_inconsistency}
\end{figure}

Figure~\ref{fig:payoff_unique_beliefs} shows the mean number of unique beliefs for the consensus operator after $50,000$ iterations plotted against the inconsistency threshold $\gamma$. For $\gamma \geq 0.5$ applying the consensus operator results in the population of agents converging on a single shared belief. Figures~\ref{fig:payoff_vagueness} and~\ref{fig:payoff_entropy} show the average vagueness and entropy of the beliefs held across the population of agents at the end of the simulation. In Figure~\ref{fig:payoff_vagueness} we see that for $\gamma \geq 0.5$ the beliefs resulting from applying the consensus operator are crisp. Figure~\ref{fig:payoff_entropy} shows that the mean entropy values decreases as $\gamma$ increases resulting in an average entropy of $0$ for $\gamma \geq 0.5$. Overall then, as in section 4, for $\gamma \geq 0.5$ the population of agents converge on a single shared belief which is both crisp and certain. Figure~\ref{fig:payoff_inconsistency} shows the average pairwise inconsistency of the population increases at lower threshold values prior to exceeding the mean inconsistency value of $\frac{2}{9}$. For inconsistency thresholds $\gamma \geq 0.3$, the average pairwise inconsistency decreases with the number of unique beliefs (seen in Figure~\ref{fig:payoff_unique_beliefs}) as the population converges towards a single shared belief.

\begin{figure}[t]
\centering
\includegraphics[width=0.45\textwidth]{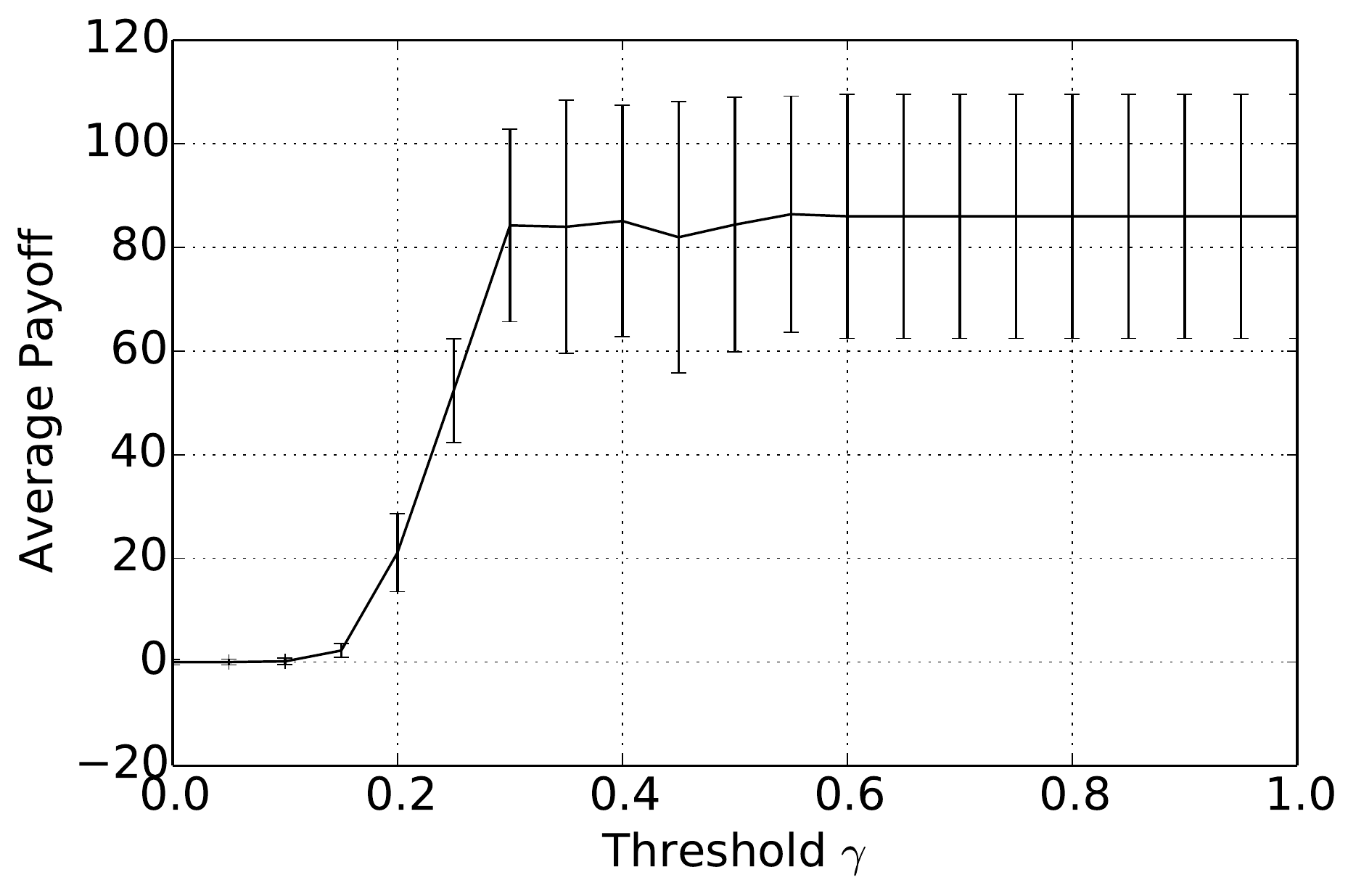}
\caption{Average payoff after $50,000$ iterations for varying inconsistency thresholds $\gamma$ and $|\mathcal{L}| = 5$. \protect 
}
\label{fig:payoff_payoff}
\end{figure}

Figure~\ref{fig:payoff_payoff} shows the average quality of beliefs (Definition~\ref{payoff}) at the end of the simulation, plotted against $\gamma$  and given as the percentage of the maximum possible quality value. For $\gamma \geq 0.5$ the consensus operator converges on a single shared crisp and certain belief with a quality value which is on average over $80\%$ of the maximum. Hence, unlike in section 4 in which convergence can be to any of the $2^n$ crisp and certain beliefs at random, agent interactions guided by relative quality converge with higher probability to those beliefs amongst the $2^n$ that are the closest to the actual state of the world. In comparison to the direct evidence scenario described in section 5 we see that the payoff shown in Figure~\ref{fig:payoff_payoff} is similar to that obtained when combining a $15\%$ direct evidence rate with consensus building based on random interactions (see Figure~\ref{fig:payoff_evidence}).

\section{Conclusions}
In this paper we have investigated consensus formation for a multi-agent system in which agents' beliefs are both vague and uncertain. For this we have adopted a formalism which combines three truth states with probability, resulting in opinions which are quantified by lower and upper belief measures. A combination operator has been introduced according to which agents are assumed to be independent and in which strictly opposing truth states are replaced with an intermediate borderline truth value. In simulation experiments we have applied this operator to random agent interactions constrained by the requirement that agreement can only be reached between agents holding beliefs which are sufficiently consistent with each other. Provided that this consistency requirement is not too restrictive then the population of agents is shown to converge on a single shared belief which is both crisp and certain. Furthermore, if combined with evidence about the state of the world, either in a direct or indirect way, then consensus building of this kind results in better convergence to the truth than just evidential belief updating alone.

Overall, these results provide some evidence for the beneficial effects of allowing agents to hold beliefs which are both vague and uncertain, in the context of consensus building. However, in this paper we have only studied pairwise interactions between agents, while in the literature it is normally intended that pooling operators should be used to aggregate beliefs across a group of agents~\cite{degroot,dietrich}. Hence, future work should extend the operator in Definition~\ref{conop} so as to allow more than two agents to reach agreement at any step of the simulation. Probabilistic pooling operators can also take account of different weights associated with the beliefs of different agents and it will be interesting to investigate if this can be incorporated in our approach. Another avenue for future research is to consider noisy evidence. Evidential updating is rarely perfect and, for example, experiments can be prone to measurement errors. An interesting question is therefore, how does the combined consensus building and updating approach described in section 5 cope with such noise? In the longer term the aim is to apply our approach to distributed decision making scenarios such as, for example, in swarm robotics.

\section*{Acknowledgements}
This research is partially funded by an EPSRC PhD studentship as part of a doctoral training partnership (grant number EP/L504919/1).

\hspace{1em}

\noindent All underlying data is included in full within this paper.

\bibliographystyle{abbrv}
\bibliography{references}

\begin{thebibliography}{10}

\bibitem{balenzuela}
P.~Balenzuela, J.~P. Pinasco, and V.~Semeshenko.
\newblock The undecided have the key: Interaction-driven opinion dynamics in a
  three state model.
\newblock In {\em PLoS ONE. 10(10)}, 2015.

\bibitem{cho}
J.-H. Cho and A.~Swami.
\newblock Dynamics of uncertain opinions in social networks.
\newblock In {\em IEEE Military Communications Conference. pp 53-64}. IEEE
  Computer Society, 2014.

\bibitem{ciucci}
D.~Ciucci, D.~Dubois, and J.~Lawry.
\newblock Borderline vs. unknown: comparing three-valued representations of
  imperfect information.
\newblock In {\em International Journal of Approximate Reasoning. 55. pp
  1866-1889}. Elsevier, 2014.

\bibitem{crosscombe}
M.~Crosscombe and J.~Lawry.
\newblock Exploiting vagueness for multi-agent consensus.
\newblock In {\em Proceedings of the Second International Workshop on Smart
  Simulations and Modelling for Complex Systems. pp 53-64}. Springer, In Press.

\bibitem{lama}
M.~S. de~la Lama, I.~G. Szendro, J.~R. Iglesias, and H.~S. Wio.
\newblock Van kampen's expansion approach in an opinion formation model.
\newblock In {\em The European Physical Journal B. 51. 435-442}. Elsevier,
  2012.

\bibitem{deffuant02}
G.~Deffuant, F.~Amblard, G.~Weisbuch, and T.~Faure.
\newblock How can extremism prevail? a study based on the relative agreement
  interaction model.
\newblock In {\em Journal of Artificial Societies and Social Simulation. 5(4)}.
  JASSS, 2002.

\bibitem{degroot}
M.~H. DeGroot.
\newblock Reaching a consensus.
\newblock In {\em Journal of the American Statistical Association. 69(345). pp
  118-121}. JSTOR, 1974.

\bibitem{dietrich}
F.~Dietrich and C.~List.
\newblock Probabilistic opinion pooling.
\newblock In {\em (Hitchcock C., Hajek A. (Eds)) Oxford Handbook of Probability
  and Philosophy}. Oxford University Press, In Press.

\bibitem{douvenkelp}
I.~Douven and C.~Kelp.
\newblock Truth approximation, social epistemology, and opinion dynamics.
\newblock In {\em Erkenntnis, Vol. 75, Issue 2, pp 271-283}. Springer, 2011.

\bibitem{duboisprade88ds}
D.~Dubois and H.~Prade.
\newblock Representation and combination of uncertainty with belief functions
  and possibility.
\newblock In {\em Computational Intelligence, Vol. 4, pp 244-264}, 1988.

\bibitem{hegselmann02}
R.~Hegselmann and U.~Krause.
\newblock Opinion dynamics and bounded confidence: Models, analysis, and
  simulation.
\newblock In {\em Journal of Artificial Societies and Social Simulation. 5(3).}
  JASSS, 2002.

\bibitem{hegselmann05}
R.~Hegselmann and U.~Krause.
\newblock Opinion dynamics driven by various ways of averaging.
\newblock In {\em Computational Economics. 25(4). pp 381-405.} Springer, 2005.

\bibitem{keefe}
R.~Keefe and P.~Smith.
\newblock {\em Vagueness: A Reader}.
\newblock MIT Press, 2002.

\bibitem{lawrydubois}
J.~Lawry and D.~Dubois.
\newblock A bipolar framework for combining beliefs about vague propositions.
\newblock In {\em Proceedings of the Thirteenth International Conference on the
  Principles of Knowledge Representation and Reasoning. pp 530-540}. AAAI,
  2012.

\bibitem{lawrytang}
J.~Lawry and Y.~Tang.
\newblock On truth-gaps, bipolar belief and the assertability of vague
  propositions.
\newblock In {\em Artificial Intelligence. 191. pp 20-41}. Elsevier, 2012.

\bibitem{list}
C.~List, C.~Elsholtz, and T.~D. Seeley.
\newblock Independence and interdependence in collective decision making: an
  agent-based model of nest-site choice by honeybee swarms.
\newblock In {\em Philosophical Transactions of the Royal Society. 364(1518).
  pp 755-762}. The Royal Society, 2009.

\bibitem{perron}
E.~Perron, D.~Vasudevan, and M.~Vojnovic.
\newblock Using three states for binary consensus on complete graphs.
\newblock In {\em Proceedings of IEEE Infocom. pp 2527-2535}. IEEE
  Communications Society, 2009.

\bibitem{reigler}
A.~Reigler and I.~Douven.
\newblock Extending the hegselmann frause model iii: From single beliefs to
  complex belief states.
\newblock In {\em Episteme, Vol. 6, pp 145-163}, 2009.

\bibitem{shafer}
G.~Shafer.
\newblock {\em A Mathematical Theory of Evidence}.
\newblock Princeton University Press, Princeton, 1st edition, 1976.

\bibitem{stone}
M.~Stone.
\newblock The opinion pool.
\newblock In {\em Annals of Mathematical Statistics. 32(4). pp 1339-1342}.
  Institute of Physics Publishing, 2004.

\bibitem{vazquez}
F.~Vazquez and S.~Redener.
\newblock Ultimate fate of constrained voters.
\newblock In {\em Journal of Physics A: Mathematical and General. 37. pp
  8479-8494}. Institute of Physics Publishing, 2004.

\end{thebibliography}

\end{document}